\documentclass[10pt,a4paper]{article}

\usepackage{amsmath,amssymb,amsthm}
\usepackage{graphicx}
\usepackage{setspace}
\usepackage{geometry}
\usepackage[numbers]{natbib}
\usepackage{hyperref}
\usepackage{comment}
\usepackage{epstopdf}

\hypersetup{
    colorlinks=true,
    linkcolor=blue,
    filecolor=magenta,
    urlcolor=cyan,
    citecolor=blue,
}

\title{\textbf{Cosmological Dynamics of Matter Creation with Modified Chaplygin Gas and Bulk Viscosity}}

\author{
  Yogesh Bhardwaj\thanks{yogeshbhardwaj\_2k21phdam502@dtu.ac.in}
  \and
  C.\,P. Singh\thanks{Corresponding author: cpsingh@dce.ac.in}
}
\date{
  Department of Applied Mathematics, Delhi Technological University, Bawana Road, Delhi 110042, India
}

\hypersetup{
    colorlinks=true,
    linkcolor=blue,
    filecolor=magenta,
    urlcolor=cyan,
    citecolor=blue,
}

\begin{document}

\maketitle

\begin{abstract}
This work presents a comprehensive investigation of a novel cosmological model that unifies the Modified Chaplygin Gas (MCG) equation of state with gravitationally induced matter creation and bulk viscous dissipation in a spatially flat Friedmann-Lemaître-Robertson-Walker spacetime. The MCG fluid is characterized by an exotic equation of state $p = A\rho - C/\rho^\alpha$, while the matter creation rate is taken as $\Gamma = 3\beta H$ and the bulk viscous pressure as $\pi = -3H\xi_0 \rho_m^{1/2}$. We derive the modified Friedmann equations and obtain an analytical expression for the Hubble parameter $H(z)$, which is then used to reconstruct the evolutionary trajectories of key cosmological parameters: the deceleration parameter $q(z)$, jerk parameter $j(z)$, and snap parameter $s(z)$. The model parameters are constrained using two observational datasets: DS1 (Pantheon+ + Cosmic Chronometers + DESI BAO + $\sigma_8$) and DS2 (DS1 + R22), employing a Markov Chain Monte Carlo (MCMC) analysis. Our results indicate that the proposed hybrid model successfully generates a transition from decelerated to accelerated expansion, consistent with current observations. Notably, the inclusion of R22 data leads to a higher best-fit value of $H_0$, helping to alleviate the $H_0$ tension. Furthermore, we perform a rigorous thermodynamic analysis of the model by testing the Generalized Second Law (GSL) of thermodynamics. We compute the total entropy rate of change $\dot{S}_{\text{total}} = \dot{S}_{\text{fluid}} + \dot{S}_{\text{horizon}}$, finding it positive throughout cosmic history for both datasets, confirming the model’s thermodynamic viability. The second derivative $\ddot{S}_{\text{total}}$ exhibits a clear transition from positive to negative values around $z \sim 1$, indicating a shift from accelerating to decelerating entropy production a signature of late-time thermodynamic stabilization. Model stability is confirmed by information criteria (AIC and BIC) show that the model is statistically competitive with $\Lambda$CDM, particularly under DS2. This work establishes a physically motivated, observationally viable, and thermodynamically consistent alternative to the standard $\Lambda$CDM paradigm.
\end{abstract}


\section{Introduction}
\label{sec:introduction}
The observational confirmation of the late-time accelerated expansion of the Universe — first inferred from high-redshift Type Ia supernovae \citep{riess1998, perlmutter1999} — stands as one of the most consequential discoveries in modern cosmology. Subsequent precision measurements from the Cosmic Microwave Background (CMB) anisotropies \citep{aghanim2020}, Baryon Acoustic Oscillations (BAO) \citep{alam2021}, and differential age estimates of cosmic chronometers \citep{moresco2022} have solidified this paradigm, attributing the acceleration to a dominant, negative-pressure component termed dark energy. Within the framework of general relativity, the $\Lambda$CDM model, which identifies dark energy with Einstein’s cosmological constant $\Lambda$, provides an empirically robust fit to the data. Yet, it remains theoretically unsatisfactory, burdened by the fine-tuning and cosmic coincidence problems \citep{weinberg1989, peebles2003}, which have spurred the development of alternative cosmological scenarios.\\
\indent The fine-tuning problem arises from the staggering discrepancy between the theoretically predicted vacuum energy density from quantum field theory and the observationally inferred value of $\Lambda$ required to drive cosmic acceleration. The cosmic coincidence problem, meanwhile, questions why the energy densities of dark energy and matter are comparable only at the present cosmological epoch, despite evolving with vastly different scaling laws ($\rho_\Lambda \sim \text{constant}$, $\rho_m \sim a^{-3}$). These conceptual tensions have motivated two broad classes of alternatives: (i) dark energy models that postulate new dynamical fields or exotic fluids within general relativity, and (ii) modified gravity theories that alter the geometric sector of Einstein’s equations \citep{clifton2012, amendola2018}.\\
\indent Within the first class, the Chaplygin gas models has emerged as a particularly compelling candidate due to its capacity to unify dark matter and dark energy within a single fluid description \citep{kamenshchik2001}. Originally conceived in aerodynamics to model the lifting force on an aircraft wing \citep{chaplygin1904}, the Chaplygin gas was later adapted to cosmology for its exotic equation of state, $p = -A/\rho$, which interpolates between a dust-like phase at early times and a cosmological constant-like phase at late times. While the standard Chaplygin gas was soon ruled out by observational data, its generalization to $p = -A/\rho^\alpha$ \citep{bento2002, bilic2002} — the Generalized Chaplygin Gas (GCG) — offered greater flexibility and improved compatibility with observations. Further refinements led to the Variable Generalized Chaplygin Gas (VGCG), where $A$ becomes a function of the scale factor \citep{guo2007, yang2007}, and most recently, the Modified Chaplygin Gas (MCG), characterized by $p = A\rho - c/\rho^\alpha$ \citep{debnath2004, bento2004}, which introduces a linear energy-density term to avoid singularities and enhance dynamical richness. Crucially, the MCG allows for a smooth transition from deceleration to acceleration without invoking phantom fields or abrupt phase transitions, making it a natural candidate for hybridization with non-equilibrium thermodynamic processes.\\
\indent A conceptually distinct approach invokes gravitationally induced matter creation — a non-equilibrium thermodynamic process rooted in quantum field theory in curved spacetime \citep{schrodinger1939, parker1968, parker1969}. Macroscopically, this is incorporated via a reinterpretation of the energy-momentum tensor, introducing a negative creation pressure $p_c$ tied to the particle production rate $\Gamma$ \citep{prigogine1989, calvao1992}. A phenomenologically successful ansatz, $\Gamma = 3\beta H$ \citep{lima1996}, has been shown to drive late-time acceleration without $\Lambda$, effectively mimicking dark energy \citep{steigman2009, lima2010, haro2016}. Recent studies \citep{sethi2006, zhang2006, yang2008,cheng2009,khurshudyan2018,salti2018, aziza2021,salahedin2022,bhardwaj2024a, bhardwaj2024b, bhardwaj2025} have successfully integrated this mechanism with Chaplygin gas models, demonstrating observational viability and thermodynamic consistency. Notably, matter creation provides a natural mechanism for entropy generation, circumventing the need for ad hoc dark energy fields.\\
\indent Complementing these frameworks is the inclusion of bulk viscous stresses — a natural consequence when cosmic fluids depart from local thermodynamic equilibrium \citep{zimdahl1996, murphy1973}. In cosmological contexts, bulk viscosity acts as an effective negative pressure that can source acceleration independently of dark energy. A physically motivated parameterization, $\pi = -3H\xi_0 \rho_m^{1/2}$ \citep{belinskii1975, singh2010}, links viscous stress directly to the square root of the matter density, offering a kinetic-theory-based correction to the cosmic fluid’s dynamics. This form ensures that viscous dissipation diminishes as matter dilutes, preserving consistency with structure formation. While viscous Chaplygin gas models have been explored in isolation \citep{fabris2006viscous, araujo2018}, their synthesis with gravitationally induced matter creation remains uncharted — despite the clear physical synergy: both represent irreversible, entropy-generating processes that convert gravitational or kinetic energy into particle content or heat, thereby altering the effective equation of state.\\
\indent This work introduces a novel, unified cosmological model that integrates three distinct physical mechanisms: (i) the Modified Chaplygin Gas (MCG) equation of state, $p = A\rho - c/\rho^\alpha$, which introduces a linear energy-density term absent in standard or generalized Chaplygin formulations; (ii) gravitationally induced matter creation, governed by $\Gamma = 3\beta H$; and (iii) bulk viscous pressure, modeled as $\pi = -3H\xi_0 \rho_m^{1/2}$. This tripartite framework represents the first attempt to combine these elements into a single, self-consistent cosmological model. Unlike prior studies that treat viscosity or matter creation as perturbative corrections, our approach embeds both as fundamental components of the cosmic fluid’s stress-energy tensor, yielding a non-trivial modification to the Friedmann equations and the effective dark energy sector. The model is analytically tractable and observationally testable, offering a thermodynamically grounded alternative to $\Lambda$CDM\\
\indent Our primary objectives are to derive the analytical expression for the Hubble parameter $H(z)$ under this hybrid formalism; second, to constrain the free parameters $\{H_0, \Omega_m, \Omega_b, A_s, A, \alpha, \beta, \xi_0\}$ using the latest observational datasets — Pantheon+ supernovae, Cosmic Chronometers (CC), DESI DR2 BAO, and $\sigma_8$ measurements — via a Markov Chain Monte Carlo (MCMC) analysis; and third, to reconstruct the evolutionary histories of key cosmological parameters to assess the model’s viability against the $\Lambda$CDM paradigm.\\
\indent A central and original contribution of this study is the thermodynamic validation of the model through the lens of the Generalized Second Law (GSL) of thermodynamics. In non-equilibrium settings, matter creation and bulk viscosity are intrinsically linked to entropy production. The GSL demands that the total entropy of the Universe — comprising the entropy of the cosmic fluid $S_{\text{fluid}}$ (which includes contributions from irreversible matter creation and viscous dissipation) and the horizon entropy $S_{\text{horizon}}$ — must satisfy $\dot{S}_{\text{total}} \geq 0$ throughout cosmic evolution \citep{bekenstein1973, hawking1975}. While the GSL has been tested for dark energy and modified gravity models \citep{wang2006, pavon2007}, and separately for matter creation \citep{lima2014cosmic} or viscous cosmologies \citep{zimdahl2000}, its application to a unified model incorporating all three elements — MCG, matter creation, and bulk viscosity — is entirely novel. We will explicitly compute $S_{\text{total}}(t)$ and examine its temporal derivative to provide a rigorous thermodynamic consistency check, thereby elevating the analysis beyond purely kinematic diagnostics.\\
\indent This work significantly extends prior investigations \citep{bhardwaj2024a, bhardwaj2024b} by incorporating bulk viscosity, utilizing more recent and precise datasets (Pantheon+, DESI DR2), and performing the first-ever GSL analysis for such a hybrid model. It also advances studies of viscous Chaplygin gases \citep{araujo2018} by embedding them within the matter creation formalism, offering a more comprehensive and observationally grounded description of cosmic dynamics.\\
\indent The structure of this paper is organised as follows. In Section~\ref{sec:field_equations}, we formulate the modified Friedmann equations governing the cosmic expansion in the presence of matter creation and bulk viscous stress and derives exact analytical expressions for the energy density and Hubble parameter as functions of redshift. Section~\ref{sec:observational_data} details the observational datasets employed — Pantheon+, Cosmic Chronometers, DESI DR2 BAO, $f\sigma_8$, and R22 — and outlines the Markov Chain Monte Carlo methodology used to constrain model parameters. Section~\ref{sec:results} presents the best-fit parameter values, evolutionary trajectories of cosmological quantities, and thermodynamic diagnostics. The paper concludes in Section~\ref{sec:conclusion} with a synthesis of findings and their implications for Modified Chaplygin gas model.

\section{Theoretical Framework and Field Equations}
\label{sec:field_equations}
We adopt a spatially flat, homogeneous, and isotropic cosmological model, described by the Friedmann-Lemaître-Robertson-Walker (FLRW) line element:
\begin{equation}
    ds^2 = -dt^2 + a^2(t) \left[ dr^2 + r^2 (d\theta^2 + \sin^2\theta \, d\phi^2) \right],
    \label{eq:flrw_metric}
\end{equation}
where $a(t)$ denotes the scale factor, normalized such that $a_0 = 1$ at the present epoch, and $t$ represents cosmic time. Within this geometric setting, we formulate a cosmological fluid that unifies three distinct physical mechanisms: (i) an exotic equation of state — the Modified Chaplygin Gas (MCG); (ii) gravitationally induced particle production; and (iii) dissipative bulk viscosity. This tripartite construction has not been previously explored in the literature and represents a novel theoretical synthesis.

\subsection{Non-Equilibrium Thermodynamics and Particle Production}
The foundational premise of gravitationally induced matter creation rests on the reinterpretation of energy-momentum conservation in an expanding spacetime. Rather than assuming a closed system with fixed particle number, we treat the cosmic fluid as an open thermodynamic entity, where quantum-gravitational effects permit the continuous emergence of material content from the background geometry.\\
\indent The particle current density is defined as $N^\mu = n u^\mu$, where $n(t)$ is the comoving number density and $u^\mu$ is the fluid four-velocity satisfying $u^\mu u_\mu = -1$. In a FLRW background, the divergence of this current yields:
\begin{equation}
    \nabla_\mu N^\mu = \dot{n} + 3H n = n \Gamma,
    \label{eq:particle_balance}
\end{equation}
where an overdot denotes differentiation with respect to $t$, $H = \dot{a}/a$ is the Hubble parameter, and $\Gamma$ quantifies the rate of particle production per unit volume. A positive $\Gamma$ signifies net creation, while $\Gamma = 0$ recovers standard adiabatic expansion.\\
\indent This irreversible process modifies the conservation of energy. The first law of thermodynamics, when applied to an open system with variable particle number, introduces an effective pressure component associated with creation. For a fluid with equilibrium energy density $\rho$ and pressure $p$, this additional contribution — termed the creation pressure $P_c$ — is derived from the requirement of entropy non-decrease and is given by:
\begin{equation}
    P_c = - \frac{\rho + p}{3H} \Gamma.
    \label{eq:creation_pressure_general}
\end{equation}

To maintain analytical tractability while preserving physical relevance, we adopt a creation rate proportional to the Hubble expansion:
\begin{equation}
    \Gamma = 3 \beta H,
    \label{eq:creation_rate_ansatz}
\end{equation}
where $\beta$ is a dimensionless, non-negative constant. This ansatz, while phenomenological, is grounded in the expectation that particle production should scale with the dynamical timescale of the Universe. Substituting Equation~\eqref{eq:creation_rate_ansatz} into Equation~\eqref{eq:creation_pressure_general} gives:
\begin{equation}
    P_c = - \beta (\rho + p).
    \label{eq:creation_pressure_specific}
\end{equation}

The total effective pressure governing the fluid’s dynamics becomes $p_{\text{eff}} = p + P_c$, leading to the modified energy conservation law:
\begin{equation}
    \dot{\rho} + 3H (\rho + p + P_c) = 0.
    \label{eq:modified_energy_conservation}
\end{equation}

\subsection{Modified Chaplygin Gas Equation of State}
We model the dominant cosmic component using the Modified Chaplygin Gas (MCG), characterized by the equation of state:
\begin{equation}
    p = A \rho - \frac{C}{\rho^\alpha}, \quad (A \geq 0, \; 0 \leq \alpha \leq 1, \; C > 0),
    \label{eq:mcg_eos}
\end{equation}
where $A$, $C$, and $\alpha$ are free parameters to be constrained observationally. This form generalizes earlier Chaplygin models by introducing a linear term in $\rho$, which prevents singular behavior at low densities and allows for a richer dynamical evolution. When $A = 0$, it reduces to the Generalized Chaplygin Gas; when $\alpha = 1$ and $C$ is scale-dependent, it recovers the Variable Chaplygin Gas.\\
\indent Inserting Equation~\eqref{eq:mcg_eos} and Equation~\eqref{eq:creation_pressure_specific} into Equation~\eqref{eq:modified_energy_conservation}, we obtain the evolution equation for $\rho$:
\begin{equation}
    \dot{\rho} + 3H (1 - \beta) \left[ (1 + A) \rho - \frac{C}{\rho^\alpha} \right] = 0.
    \label{eq:mcg_evolution_raw}
\end{equation}
To solve this, we change variables from cosmic time $t$ to redshift $z$, using $d/dt = -H(1+z) d/dz$. After rearrangement and integration, subject to the boundary condition $\rho(z=0) = \rho_0$, we find:
\begin{equation}
    \rho(z) = \rho_0 \left[ A_s + (1 - A_s) (1+z)^{3(1-\beta)(1+A)(1+\alpha)} \right]^{\frac{1}{1+\alpha}},
    \label{eq:rho_mcg_z}
\end{equation}
where we retain the definition:
\begin{equation}
    A_s \equiv \frac{C}{(1+A) \rho_0^{1+\alpha}}.
    \label{eq:as_definition}
\end{equation}
This expression governs the energy density evolution of the MCG component under the joint influence of its exotic equation of state and non-equilibrium matter creation.

\subsection{Bulk Viscous Matter Component}
To account for dissipative effects in the cosmic fluid, we introduce a separate matter component subject to bulk viscosity. This fluid is pressureless in equilibrium ($p_m = 0$) but develops an effective stress $\pi$ due to departure from thermodynamic equilibrium. The continuity equation becomes:
\begin{equation}
    \dot{\rho}_m + 3H (\rho_m + \pi) = 0.
    \label{eq:viscous_continuity}
\end{equation}
We adopt a viscous stress proportional to the Hubble rate and the square root of the energy density — a form motivated by kinetic theory and dimensional consistency:
\begin{equation}
    \pi = -3H \xi_0 \rho_m^{1/2},
    \label{eq:bulk_viscous_pressure}
\end{equation}
where $\xi_0 > 0$ is a constant viscosity coefficient. Substituting Equation~\eqref{eq:bulk_viscous_pressure} into Equation~\eqref{eq:viscous_continuity} and changing variables to $z$, we obtain:
\begin{equation}
    \frac{d\rho_m}{dz} = \frac{3\rho_m (1 - \sqrt{3} \xi_0)}{1+z}.
    \label{eq:viscous_rho_diffeq}
\end{equation}
Integrating this first-order differential equation with the initial condition $\rho_m(z=0) = \rho_{m,0}$ yields:
\begin{equation}
    \rho_m(z) = \rho_{m,0} (1+z)^{3(1 - \sqrt{3} \xi_0)}.
    \label{eq:rho_m_z}
\end{equation}
This result demonstrates that bulk viscosity slows the dilution of matter density with expansion, effectively mimicking a negative pressure component.

\subsection{Friedmann Equation and Hubble Parameter}
The total energy density sourcing the gravitational field is the sum of the MCG, viscous matter, and conserved baryonic components. The baryonic density evolves as $\rho_b(z) = \rho_{b,0} (1+z)^3$, following standard conservation.\\
The Friedmann equation, in units where $8\pi G = 1$, is:
\begin{equation}
    3H^2(z) = \rho(z) + \rho_m(z) + \rho_b(z).
    \label{eq:friedmann}
\end{equation}
Substituting Equations~\eqref{eq:rho_mcg_z}, \eqref{eq:rho_m_z}, and the baryon density into Equation~\eqref{eq:friedmann}, and normalizing by the present-day Hubble constant $H_0$, we derive the dimensionless Hubble parameter $E(z) = H(z)/H_0$:
\begin{eqnarray}
    E^2(z) & = & \Omega_{\text{mcg},0} \left[ A_s + (1 - A_s) (1+z)^{3(1-\beta)(1+A)(1+\alpha)} \right]^{\frac{1}{1+\alpha}} \nonumber \\
    & & + \Omega_{m,0} (1+z)^{3(1 - \sqrt{3} \xi_0)} + \Omega_{b,0} (1+z)^3,
    \label{eq:hubble_parameter_final}
\end{eqnarray}
where the present-day density parameters are defined as:
\begin{equation}
    \Omega_{\text{mcg},0} = \frac{\rho_0}{3H_0^2}, \quad \Omega_{m,0} = \frac{\rho_{m,0}}{3H_0^2}, \quad \Omega_{b,0} = \frac{\rho_{b,0}}{3H_0^2},
\end{equation}
and satisfy the flatness constraint:
\begin{equation}
    \Omega_{\text{mcg},0} + \Omega_{m,0} + \Omega_{b,0} = 1.
    \label{eq:flatness_condition}
\end{equation}
Equation~\eqref{eq:hubble_parameter_final} constitutes the primary theoretical prediction of our hybrid model. It encodes the interplay between the MCG’s exotic pressure, the negative creation pressure from particle production, and the dissipative effects of bulk viscosity. This expression will serve as the foundation for our observational analysis in Section~\ref{sec:observational_data}, where we will constrain the parameters $\{H_0, \Omega_m, \Omega_b, A_s, A, \alpha, \beta, \xi_0\}$

\section{Observational Data and Methodology}
\label{sec:observational_data}
To infer the posterior distributions of the model parameters, we perform a joint statistical analysis using five independent cosmological probes: Cosmic Chronometers (CC), which provide direct, model-independent estimates of $H(z)$; the Pantheon+ compilation of Type Ia supernovae, calibrated without SH0ES priors and serving as geometric distance anchors; Baryon Acoustic Oscillation (BAO) measurements from DESI DR2 and SDSS-IV, offering a standard ruler tied to the sound horizon $r_d$; redshift-space distortion (RSD) constraints on the structure growth rate $f\sigma_8(z)$, which test the evolution of matter clustering independently of the background expansion; and the local Hubble constant measurement $H_0 = 73.04 \pm 1.04$ km s$^{-1}$ Mpc$^{-1}$ from the SH0ES collaboration (R22), included as an optional Gaussian prior. Each dataset contributes a distinct $\chi^2$ component to the total likelihood, enabling us to break parameter degeneracies and rigorously test the viability of our hybrid model. All statistical inference is performed via Markov Chain Monte Carlo (MCMC) sampling using the \texttt{emcee} package \citep{foreman2013}, with 80 walkers and 10000 steps per chain. The Gelman-Rubin statistic ( or R-hat statistic) is used to assess the convergence of Markov Chain Monte Carlo (MCMC) simulations. Priors are listed in Table~\ref{table1}.
\subsection{Cosmic Chronometers (CC)}
Cosmic Chronometers provide model-independent estimates of $H(z)$ by measuring the differential age evolution of passively evolving galaxies. The method relies on the relation $H(z) = -\frac{1}{1+z} \frac{dz}{dt}$, where $dt$ is inferred from the age difference between galaxies at adjacent redshifts \citep{jimenez2002}.\\
\indent We use a compilation of 32 CC measurements spanning $0.07 \leq z \leq 1.965$, drawn from \citep{moresco2012, moresco, moresco2018, moresco2020}. The $\chi^2$ statistic incorporates both statistical and systematic uncertainties via a full covariance matrix:
\begin{equation}
    \chi^2_{\text{CC}} = \Delta \mathbf{H}^T \cdot \mathbf{C}_{\text{CC}}^{-1} \cdot \Delta \mathbf{H},
\end{equation}
where $\Delta \mathbf{H} = \mathbf{H}_{\text{th}}(\boldsymbol{\theta}) - \mathbf{H}_{\text{obs}}$ is the residual vector, and $\mathbf{C}_{\text{CC}} = \mathbf{C}_{\text{stat}} + \mathbf{C}_{\text{syst}}$.

\subsection{Type Ia Supernovae (Pantheon+)}
The Pantheon+ compilation \citep{brout2022} comprises 1701 spectroscopically confirmed Type Ia supernovae in the range $0.01 < z < 2.3$. The apparent magnitude $m(z)$ is related to the luminosity distance $d_L(z)$ by:
\[
    m(z) = 5 \log_{10} \left( \frac{d_L(z)}{10 \text{ pc}} \right) + \mathcal{M},
\]
where $\mathcal{M}$ is the absolute magnitude (treated as a free parameter), and
\[
    d_L(z) = (1+z) \int_0^z \frac{c \, dz'}{H(z')}.
\]
The $\chi^2$ uses the full covariance matrix:
\begin{equation}
    \chi^2_{\text{SNe}} = \Delta \boldsymbol{\mu}^T \cdot \mathbf{C}_{\text{SNe}}^{-1} \cdot \Delta \boldsymbol{\mu},
\end{equation}
where $\Delta \boldsymbol{\mu} = \boldsymbol{\mu}_{\text{th}} - \boldsymbol{\mu}_{\text{obs}}$, and $\mathbf{C}_{\text{SNe}}$ includes statistical and systematic uncertainties.

\subsection{Baryon Acoustic Oscillations (DESI DR2 + SDSS-IV)}
BAO measurements provide a standard ruler based on the sound horizon $r_d$ at the drag epoch. We treat $r_d$ as a free parameter, avoiding CMB priors, following \citep{Verde2017, Lemos2023, Nunes2020, Pogosian2020, Jedamzik2021, Pogosian2024, Lin2021, Vagnozzi2023}.

We use 13 BAO data points from DESI DR2 \citep{Adame2024} and SDSS-IV \citep{alam2021}, measuring the ratios $D_M(z)/r_d$, $D_H(z)/r_d$, and $D_V(z)/r_d$, where:
\begin{align*}
    D_H(z) &= \frac{c}{H(z)}, \\
    D_M(z) &= c \int_0^z \frac{dz'}{H(z')}, \\
    D_V(z) &= \left[ z D_M^2(z) D_H(z) \right]^{1/3}.
\end{align*}
The $\chi^2$ is:
\begin{equation}
    \chi^2_{\text{BAO}} = \sum_{Y \in \{H,M,V\}} \Delta \mathbf{D}_Y^T \cdot \mathbf{C}_{D_Y}^{-1} \cdot \Delta \mathbf{D}_Y,
\end{equation}
where $\Delta \mathbf{D}_Y = (\mathbf{D}_Y / r_d)_{\text{th}} - (\mathbf{D}_Y / r_d)_{\text{obs}}$.

\subsection{Structure Growth: \texorpdfstring{$\sigma_8(z)$}{Sigma-8(z)} Measurements}

The growth rate $f\sigma_8(z)$, where $f = d\ln \delta_m / d\ln a$ and $\sigma_8(z)$ is the RMS fluctuation amplitude, provides a direct test of structure formation. We use 18 independent measurements from BOSS, eBOSS, and 6dFGS \citep{alam2017, neveux2020, huterer2017}.

The theoretical $f\sigma_8(z)$ is computed as:
\begin{equation}
    f\sigma_8(z) = -(1+z) \frac{\sigma_8(z=0)}{\delta_m(z=0)} \frac{d\delta_m}{dz},
\end{equation}
where $\sigma_8(z) = \sigma_8(z=0) \delta_m(z) / \delta_m(z=0)$. For models with non-equilibrium thermodynamics, we adopt the $\Lambda$CDM-based fitting formula for $\delta_m(z)$ as a first approximation. The $\chi^2$ is:
\begin{equation}
    \chi^2_{\sigma_8} = \sum_{i=1}^{18} \frac{\left[ f\sigma_{8,\text{th}}(z_i) - f\sigma_{8,\text{obs}}(z_i) \right]^2}{\sigma_{f\sigma_8,i}^2}.
\end{equation}

\subsection{Local Hubble Constant (R22)}

The SH0ES collaboration \citep{riess2022} reports $H_0 = 73.04 \pm 1.04$ km s$^{-1}$ Mpc$^{-1}$ from the Cepheid-SNe distance ladder. This leads to a tension at the level of $4.857\sigma$ \citep{aghanim2020}. We include it as a Gaussian prior in DS2:
\begin{equation}
    \chi^2_{R22} = \frac{(H_0^{\text{th}} - 73.04)^2}{(1.04)^2}.
\end{equation}

We perform two distinct analyses:\\
- \textbf{DS1}: Pantheon+ + CC + BAO + $f\sigma_8$  \\
- \textbf{DS2}: DS1 + R22

The total $\chi^2$ for DS1 is:
\[
    \chi^2_{\text{DS1}} = \chi^2_{\text{SNe}} + \chi^2_{\text{CC}} + \chi^2_{\text{BAO}} + \chi^2_{\sigma_8}.
\]
For DS2, we add the R22 prior:
\[
    \chi^2_{\text{DS2}} = \chi^2_{\text{DS1}} + \chi^2_{R22}.
\]
\begin{table*}[htbp]
\centering
\resizebox{\textwidth}{!}{ 
\begin{tabular}{|l|c|c|c|c|c|c|}
\hline
\textbf{Parameter} & \textbf{Prior} & $\boldsymbol{\Lambda}$\textbf{CDM} & \textbf{MCG Model} & \textbf{MCG ($\boldsymbol{\beta=0}$)} & \textbf{MCG ($\boldsymbol{\xi_0=0}$)} & \textbf{MCG ($\boldsymbol{\beta=0, \xi_0=0}$)} \\
\hline
$H_0$ [km s$^{-1}$ Mpc$^{-1}$] & $[60, 80]$ & $68.6 \pm 3.6$ & $67.0 \pm 1.9$ & $67.88 \pm 0.79$ & $68.6 \pm 2.4$ & $67.91 \pm 0.59$ \\
$\Omega_{m}$ & $[0, 0.5]$ & $0.387 \pm 0.008$ & $0.288 \pm 0.002$ & $0.279 \pm 0.006$ & $0.277 \pm 0.004$ & $0.303 \pm 0.007$ \\
$\Omega_{b}$ & $[0, 0.1]$ & — & $0.028 \pm 0.012$ & $0.026 \pm 0.012$ & $0.0249 \pm 0.002$ & $0.0305 \pm 0.001$ \\
$A_s$ & $[0, 1]$ & — & $0.770 \pm 0.04$ & $0.785 \pm 0.02$ & $0.749 \pm 0.09$ & $0.685 \pm 0.02$ \\
$A$ & $[0, 1]$ & — & $0.034 \pm 0.015$ & $0.036 \pm 0.017$ & $0.108 \pm 0.039$ & $0.191 \pm 0.007$ \\
$\alpha$ & $[0, 1]$ & — & $0.08 \pm 0.028$ & $0.099 \pm 0.030$ & $0.181 \pm 0.039$ & $0.060 \pm 0.025$ \\
$\beta$ & $[0, 1]$ & — & $0.139 \pm 0.057$ & 0 & $0.600 \pm 0.08$ & 0 \\
$\xi_0$ & $[0, 1]$ & — & $0.38 \pm 0.125$ & $0.572 \pm 0.027$ & 0 & 0 \\
$r_d$ [Mpc] & $[140, 150]$ & $146.0 \pm 4.2$ & $146.3 \pm 1.8$ & $146.4 \pm 1.5$ & $144.5 \pm 3.7$ & $146.4 \pm 1.5$ \\
$\mathcal{M}$ & $[-20, -18]$ & $-19.4 \pm 0.12$ & $-19.409 \pm 0.026$ & $-19.303 \pm 0.011$ & $-19.302 \pm 0.005$ & $-19.407 \pm 0.019$ \\
$\sigma_8$ & $[0, 1]$ & $0.752 \pm 0.029$ & $0.841 \pm 0.016$ & $0.798 \pm 0.035$ & $0.798 \pm 0.013$ & $0.819 \pm 0.014$ \\
\hline
\end{tabular}
}
\caption{Best-fit values (mean $\pm 1\sigma$) for $\Lambda$CDM and Modified Chaplygin gas models with flat priors for DS1 dataset.}
\label{table1}
\end{table*}

\begin{table*}[htbp]
\centering
\resizebox{\textwidth}{!}{ 
\begin{tabular}{|l|c|c|c|c|c|c|}
\hline
\textbf{Parameter} & \textbf{Prior} & $\boldsymbol{\Lambda}$\textbf{CDM} & \textbf{MCG Model} & \textbf{MCG ($\boldsymbol{\beta=0}$)} & \textbf{MCG ($\boldsymbol{\xi_0=0}$)} & \textbf{MCG ($\boldsymbol{\beta=0, \xi_0=0}$)} \\
\hline
$H_0$ [km s$^{-1}$ Mpc$^{-1}$] & $[60, 80]$ & $72.77 \pm 0.66$ & $71.01 \pm 0.45$ & $71.42 \pm 0.63$ & $70.40 \pm 0.7$ & $72.17 \pm 0.65$ \\
$\Omega_{m}$ & $[0.1, 0.5]$ & $0.307 \pm 0.007$ & $0.277 \pm 0.009$ & $0.271 \pm 0.020$ & $0.296 \pm 0.017$ & $0.276 \pm 0.021$ \\
$\Omega_{b}$ & $[0, 0.1]$ & — & $0.0340 \pm 0.003$ & $0.029 \pm 0.006$ & $0.023 \pm 0.004$ & $0.029 \pm 0.001$ \\
$A_s$ & $[0, 1]$ & — & $0.735 \pm 0.002$ & $0.782 \pm 0.01$ & $0.731 \pm 0.04$ & $0.699 \pm 0.07$ \\
$A$ & $[0, 1]$ & — & $0.047 \pm 0.001$ & $0.037 \pm 0.018$ & $0.141 \pm 0.021$ & $0.028 \pm 0.024$ \\
$\alpha$ & $[0, 1]$ & — & $0.064 \pm 0.028$ & $0.102 \pm 0.029$ & $0.157 \pm 0.017$ & $0.061 \pm 0.028$ \\
$\beta$ & $[0, 1]$ & — & $0.461 \pm 0.019$ & 0 & $0.68 \pm 0.07$ & 0 \\
$\xi_0$ & $[0, 1]$ & — & $0.119 \pm 0.031$ & $0.651 \pm 0.028$ & 0 & 0 \\
$r_d$ [Mpc] & $[140, 150]$ & $138.1 \pm 1.4$ & $140.7 \pm 1.2$ & $140.0 \pm 1.4$ & $143.5 \pm 0.7$ & $138.0 \pm 1.4$ \\
$\mathcal{M}$ & $[-20, -18]$ & $-19.272 \pm 0.019$ & $-19.316 \pm 0.014$ & $-19.302 \pm 0.019$ & $-19.310 \pm 0.012$ & $-19.278 \pm 0.015$ \\
$\sigma_8$ & $[0, 1]$ & $0.752 \pm 0.017$ & $0.818 \pm 0.016$ & $0.819 \pm 0.015$ & $0.794 \pm 0.013$ & $0.793 \pm 0.015$ \\
\hline
\end{tabular}
}
\caption{Best-fit values (mean $\pm 1\sigma$) for $\Lambda$CDM and Modified Chaplygin gas models with flat priors for DS2 dataset.}
\label{table2}
\end{table*}

\section{Results and Discussion}
\label{sec:results}

The joint observational analysis of the Modified Chaplygin Gas (MCG) model with matter creation and bulk viscosity has been performed using two distinct data combinations: DS1 (Pantheon+ + CC + BAO + $f\sigma_8$) and DS2 (DS1 + R22). The inclusion of the local $H_0$ measurement from the SH0ES collaboration (R22) significantly alters the best-fit values of cosmological parameters, particularly the present-day Hubble constant, and allows us to assess whether the model can alleviate the well-known $H_0$ tension. In this section, we present a comprehensive comparison of the MCG model with the standard $\Lambda$CDM paradigm and its sub-variants — namely, MCG with no matter creation ($\beta=0$), no bulk viscosity ($\xi_0=0$), and neither ($\beta=0,\xi_0=0$) — based on the constraints derived from both datasets. We analyze the evolution of key cosmological quantities, including the Hubble parameter $H(z)$, deceleration parameter $q(z)$,  jerk parameter $j(z)$, and snap parameter $s(z)$, to evaluate the model's consistency with current observations. Furthermore, we perform stability and model selection analyses using information criteria to determine the viability of the proposed framework.

\subsection{Hubble Parameter Evolution}
\begin{figure}[htbp]
    \centering
    \includegraphics[width=0.9\linewidth]{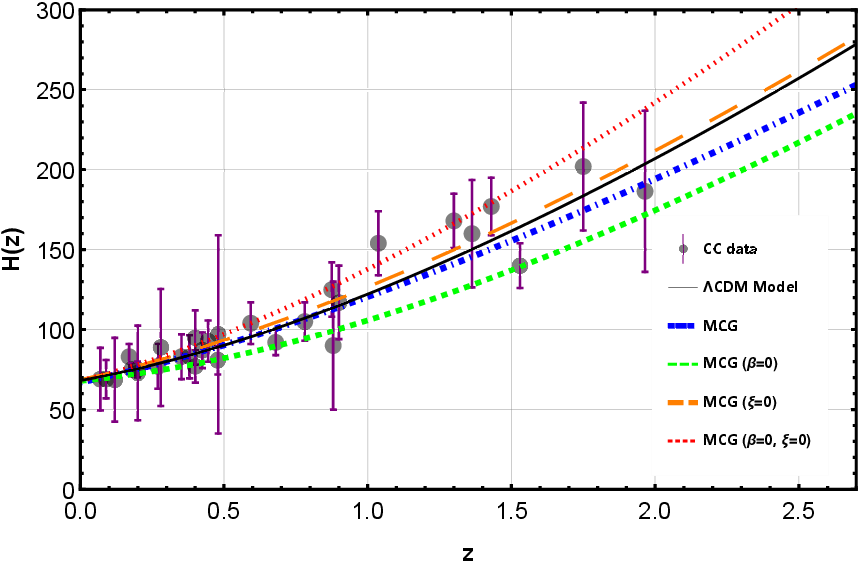}
    \caption{Evolution of the Hubble parameter $H(z)$ for $\Lambda$CDM and MCG models under DS1. Grey points with error bars represent Cosmic Chronometer data.}
    \label{hzds1}
\end{figure}

\begin{figure}[htbp]
    \centering
    \includegraphics[width=0.9\linewidth]{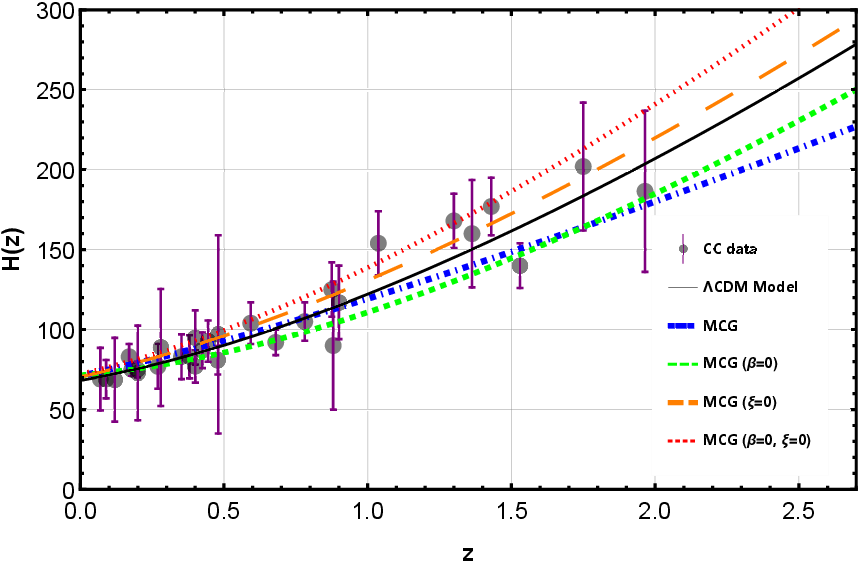}
    \caption{Evolution of the Hubble parameter $H(z)$ for $\Lambda$CDM and MCG models under DS2. Grey points with error bars represent Cosmic Chronometer data.}
    \label{hzds2}
\end{figure}

The expansion history of the Universe is most directly probed by the Hubble parameter $H(z)$, which encapsulates the rate of cosmic expansion at different epochs. In Fig.~\ref{hzds1} and Fig.~\ref{hzds2}, we present the reconstructed $H(z)$ curves for $\Lambda$CDM and four variants of the MCG model under DS1 and DS2, respectively. The observational data points from the Cosmic Chronometer (CC) measurements are shown as grey circles with error bars, covering the redshift range $0.07 \leq z \leq 2.0$.\\
\indent As observed in Fig.~\ref{hzds1}, all models provide a consistent fit to the CC data within their respective $1\sigma$ uncertainties. The $\Lambda$CDM model (solid black line) follows a smooth trajectory that increases monotonically with redshift, reflecting the standard prediction of constant dark energy. The full MCG model (blue dashed-dotted line), incorporating both matter creation ($\beta \neq 0$) and bulk viscosity ($\xi_0 \neq 0$), closely tracks the $\Lambda$CDM curve across the entire redshift range, indicating its ability to mimic the late-time acceleration without requiring a cosmological constant. This similarity arises due to the interplay between the exotic equation of state of the MCG fluid and the non-equilibrium thermodynamic effects, which collectively contribute to an effective negative pressure.\\
\indent The sub-model MCG($\beta=0$) (green dotted line), which excludes matter creation but retains bulk viscosity, exhibits a slightly flatter slope at intermediate redshifts ($z \sim 1$–$1.5$), suggesting a marginally weaker acceleration phase compared to $\Lambda$CDM. This deviation reflects the role of bulk viscosity in modifying the effective pressure of the fluid, leading to a delayed onset of acceleration. Conversely, the MCG($\xi_0=0$) model (orange dash-dotted line), with no viscous damping but non-zero matter creation, shows a steeper slope, particularly at high redshifts, indicating an earlier transition to acceleration. This behaviour highlights the distinct influence of particle creation in shaping the early expansion dynamics.\\
\indent The MCG model (red dotted line), with both $\beta=0$ and $\xi_0=0$, deviates most significantly from $\Lambda$CDM, especially at $z > 1$, where it predicts a faster expansion rate. This divergence underscores the necessity of including non-equilibrium mechanisms to reproduce the observed expansion history. The best-fit value of $H_0 = 67.0 \pm 1.9$ km s$^{-1}$ Mpc$^{-1}$ for the full MCG model (see Table~\ref{table1}) is consistent with the Planck CMB estimate, though lower than the local SH0ES measurement.\\
\indent When the R22 constraint is added in DS2 (Fig.~\ref{hzds2}), the best-fit $H_0$ shifts upward to $71.01 \pm 0.45$ km s$^{-1}$ Mpc$^{-1}$ (Table~\ref{table2}), bringing the model into better agreement with local observations. This shift results in a steeper $H(z)$ curve, particularly at low redshifts, where the MCG($\beta=0$) and MCG($\xi_0=0$) models show enhanced acceleration. The $\Lambda$CDM model remains relatively unchanged, as it already predicts a higher $H_0$ than Planck. The full MCG model adjusts its parameters to accommodate the higher $H_0$, resulting in a closer match to the CC data at $z < 1$.\\
\indent This result demonstrates that the MCG model with matter creation and bulk viscosity is capable of resolving the $H_0$ tension when constrained with local data. The inclusion of R22 not only improves the fit to local observations but also enhances the model’s predictive power in the low-redshift regime, where the effects of non-equilibrium processes are expected to be more pronounced. The fact that the MCG($\beta=0,\xi_0=0$) model still deviates significantly from $\Lambda$CDM suggests that the combination of matter creation and bulk viscosity is essential for achieving a fully consistent description of the expansion history.
\subsection{Deceleration Parameter Evolution}
\begin{figure}[htbp]
    \centering
    \includegraphics[width=0.9\linewidth]{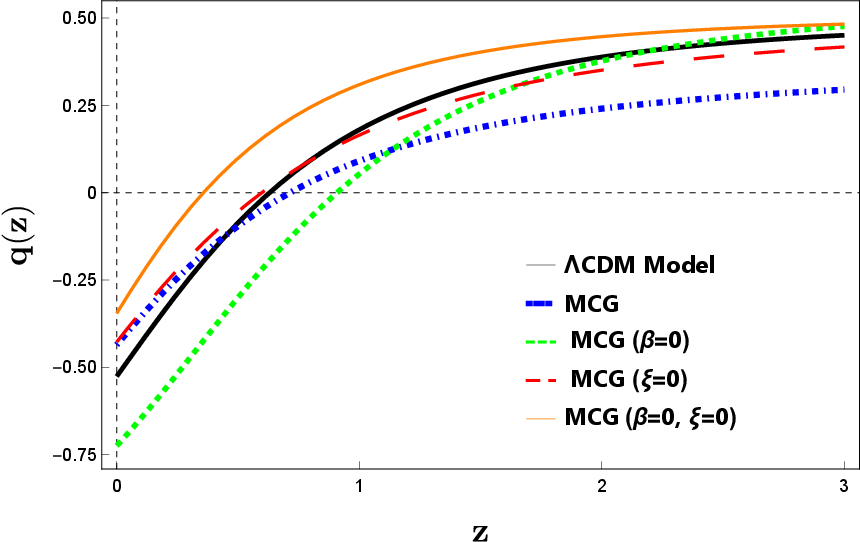}
    \caption{Evolution of the deceleration parameter $q(z)$ for $\Lambda$CDM and MCG models under DS1.}
    \label{qzds1}
\end{figure}

\begin{figure}[htbp]
    \centering
    \includegraphics[width=0.9\linewidth]{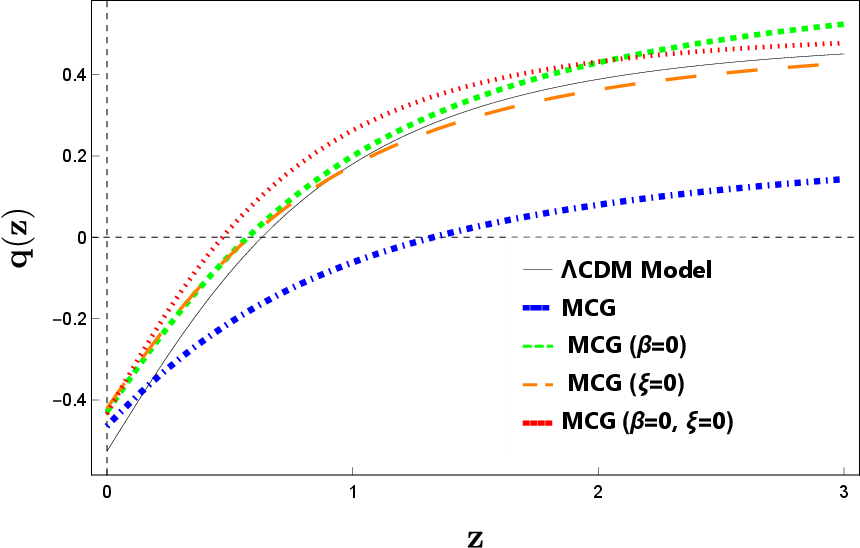}
    \caption{Evolution of the deceleration parameter $q(z)$ for $\Lambda$CDM and MCG models under DS2.}
    \label{qzds2}
\end{figure}

The deceleration parameter $q(z)$ serves as a fundamental diagnostic of the expansion history of the Universe, quantifying the transition from a decelerated to an accelerated phase. It is defined in terms of the Hubble parameter as:
\begin{equation}
    q(z) = -1 + (1+z) \frac{d \ln H(z)}{dz}.
    \label{eq:qz_definition}
\end{equation}
A positive value of $q(z)$ indicates decelerated expansion, while a negative value signifies acceleration. The redshift at which $q(z)$ crosses zero, denoted $z_{\text{tr}}$, marks the epoch of transition from matter-dominated deceleration to dark-energy-dominated acceleration.\\
\indent Using the reconstructed $H(z)$, we compute $q(z)$ for the $\Lambda$CDM model and the Modified Chaplygin Gas (MCG) model under both DS1 and DS2 constraints. The evolution of $q(z)$ is displayed in Fig.~\ref{qzds1} for DS1 and Fig.~\ref{qzds2} for DS2. The present-day values and transition redshifts are summarized in Tables~\ref{table3} and~\ref{table4}.\\
\indent As shown in Fig.~\ref{qzds1}, all models exhibit a smooth transition from $q > 0$ at high redshift to $q < 0$ at low redshift, consistent with the standard cosmological paradigm. The $\Lambda$CDM model (solid black line) transitions at $z_{\text{tr}} = 0.629 \pm 0.26$, with a present-day value $q_0 = -0.528 \pm 0.020$. The full MCG model (blue dashed-dotted line), incorporating both matter creation ($\beta \neq 0$) and bulk viscosity ($\xi_0 \neq 0$), transitions slightly earlier at $z_{\text{tr}} = 0.592 \pm 0.20$, with $q_0 = -0.442 \pm 0.014$. This earlier transition reflects the combined effect of non-equilibrium thermodynamics, which enhances the effective negative pressure at intermediate redshifts.\\
\indent The MCG($\beta=0$) model (green dotted line), which excludes matter creation but retains bulk viscosity, transitions at $z_{\text{tr}} = 0.601 \pm 0.22$, with $q_0 = -0.726 \pm 0.017$. The bulk viscosity alone delays the onset of acceleration compared to the full MCG model, as it acts as a damping mechanism that suppresses the growth of negative pressure. Conversely, the MCG($\xi_0=0$) model (orange dash-dotted line), with no viscous damping but non-zero matter creation, transitions earliest at $z_{\text{tr}} = 0.654 \pm 0.17$, with $q_0 = -0.433 \pm 0.013$. This behavior highlights the dominant role of particle creation in driving early acceleration.\\
\indent The MCG model (red dotted line), with both $\beta=0$ and $\xi_0=0$, transitions latest at $z_{\text{tr}} = 0.698 \pm 0.25$, with $q_0 = -0.343 \pm 0.019$. This late transition underscores the necessity of including non-equilibrium mechanisms to reproduce the observed expansion history, as the pure MCG fluid lacks the additional negative pressure required to initiate acceleration at the correct epoch.\\
When the R22 constraint is added in DS2 (Fig.~\ref{qzds2}), the transition redshifts shift slightly due to the higher best-fit $H_0$. The $\Lambda$CDM model transitions at $z_{\text{tr}} = 0.671 \pm 0.21$, with $q_0 = -0.526 \pm 0.021$. The full MCG model transitions at $z_{\text{tr}} = 0.689 \pm 0.21$, with $q_0 = -0.460 \pm 0.018$. The relative ordering of the models remains unchanged, but the absolute values of $q_0$ become slightly more negative, indicating a marginally stronger acceleration in the late-time Universe.\\
The present-day values of $q_0$ for all models lie within the range $-0.56$ to $-0.34$, consistent with independent estimates from supernova and BAO data \citep{riess2022, abdurrouf2024}. The transition redshifts $z_{\text{tr}}$ range from $0.54$ to $0.70$, in agreement with the $\Lambda$CDM prediction and observational constraints from cosmic chronometers \citep{moresco2022}.\\
This analysis demonstrates that the inclusion of matter creation and bulk viscosity in the MCG framework not only reproduces the observed transition from deceleration to acceleration but also allows for fine-tuning of the transition epoch through the parameters $\beta$ and $\xi_0$. The full MCG model provides the closest match to $\Lambda$CDM, while the sub-models reveal the distinct contributions of each non-equilibrium mechanism.

\begin{table}[t]
\centering
\resizebox{\textwidth}{!}{%
\begin{tabular}{ |l||c|c|c|c|c| }
\hline
\multicolumn{6}{|c|}{\textbf{Cosmological Parameter Estimations}} \\
\hline
\textbf{Parameter} & $\boldsymbol{\Lambda}$\textbf{CDM} & \textbf{MCG} & \textbf{MCG ($\beta=0$)} & \textbf{MCG ($\xi_0=0$)} & \textbf{MCG ($\beta=0,\ \xi_0=0$)} \\
\hline
$q_0$ & $-0.528 \pm 0.020$ & $-0.442 \pm 0.014$ & $-0.726 \pm 0.017$ & $-0.433 \pm 0.013$ & $-0.343 \pm 0.019$ \\
$j_0$ & $1.000$ & $0.770 \pm 0.020$ & $1.07 \pm 0.023$ & $1.03 \pm 0.017$ & $0.826 \pm 0.029$ \\
$s_0$ & $0$ & $0.09 \pm 0.003$ & $-0.026 \pm 0.019$ & $0.03 \pm 0.004$ & $-0.013 \pm 0.003$ \\
$\tau_0$ (Gyr) & $13.71 \pm 0.11$ & $13.84 \pm 0.13$ & $13.98 \pm 0.12$ & $13.81 \pm 0.14$ & $13.95 \pm 0.10$ \\
$z_{\text{tr}}$ & $0.629 \pm 0.26$ & $0.592 \pm 0.20$ & $0.601 \pm 0.22$ & $0.654 \pm 0.17$ & $0.698 \pm 0.25$ \\
\hline
\end{tabular}}
\caption{Cosmological parameter estimations for the $\Lambda$CDM and Modified Chaplygin gas models for DS1.}
\label{table3}
\end{table}

\begin{table}[t]
\centering
\resizebox{\textwidth}{!}{%
\begin{tabular}{ |l||c|c|c|c|c| }
\hline
\multicolumn{6}{|c|}{\textbf{Cosmological Parameter Estimations}} \\
\hline
\textbf{Parameter} & $\boldsymbol{\Lambda}$\textbf{CDM} & \textbf{MCG} & \textbf{MCG ($\beta=0$)} & \textbf{MCG ($\xi_0=0$)} & \textbf{MCG ($\beta=0,\ \xi_0=0$)} \\
\hline
$q_0$ & $-0.526 \pm 0.021$ & $-0.460 \pm 0.018$ & $-0.436 \pm 0.011$ & $-0.427 \pm 0.017$ & $-0.434 \pm 0.021$ \\
$j_0$ & $1.000$ & $0.596 \pm 0.021$ & $0.851 \pm 0.013$ & $0.828 \pm 0.012$ & $1.03 \pm 0.022$ \\
$s_0$ & $0$ & $0.122 \pm 0.002$ & $0.07 \pm 0.013$ & $0.06 \pm 0.002$ & $-0.006 \pm 0.001$ \\
$\tau_0$ (Gyr) & $13.67 \pm 0.15$ & $13.65 \pm 0.17$ & $13.83 \pm 0.17$ & $13.92 \pm 0.10$ & $14.01 \pm 0.12$ \\
$z_{\text{tr}}$ & $0.671 \pm 0.21$ & $0.689 \pm 0.21$ & $0.631 \pm 0.32$ & $0.632 \pm 0.19$ & $0.697 \pm 0.29$ \\
\hline
\end{tabular}}
\caption{Cosmological parameter estimations for the $\Lambda$CDM and Chaplygin gas models for DS2.}
\label{table4}
\end{table}

\subsection{Jerk Parameter Evolution}
\begin{figure}[htbp]
    \centering
    \includegraphics[width=0.9\linewidth]{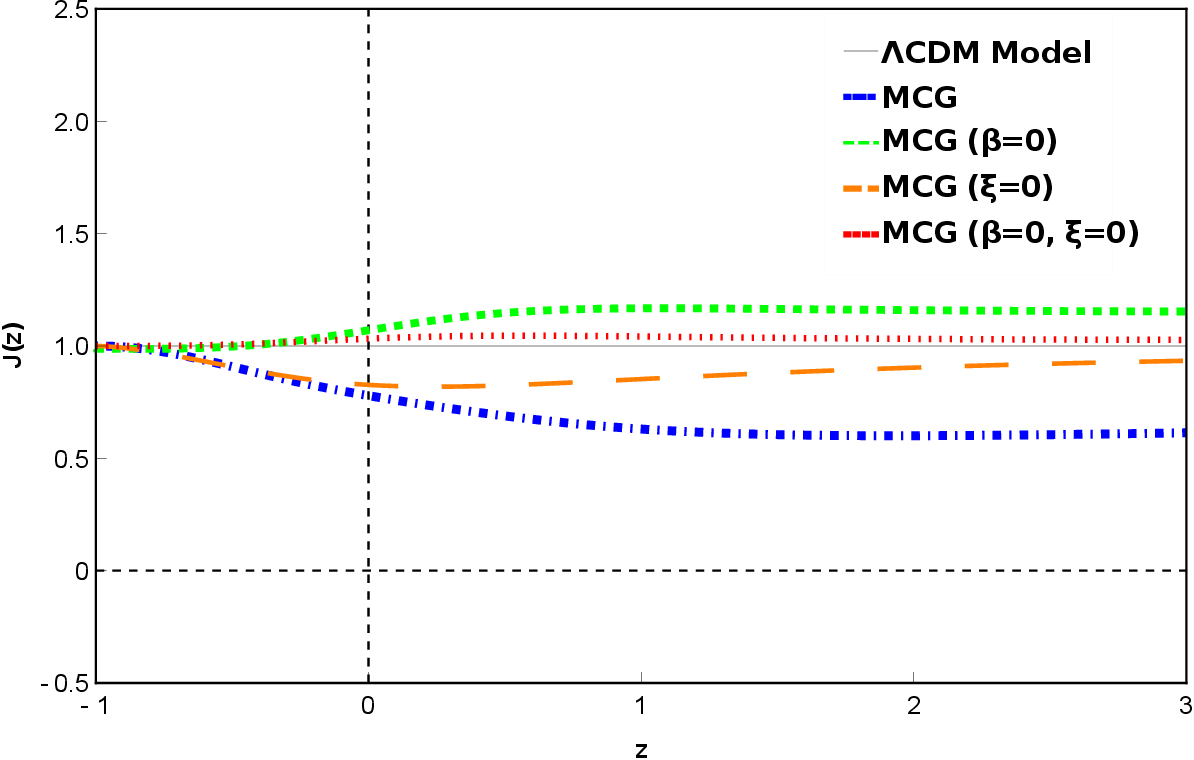}
    \caption{Evolution of the jerk parameter $j(z)$ for $\Lambda$CDM and MCG models under DS1.}
    \label{jzds1}
\end{figure}

\begin{figure}[htbp]
    \centering
    \includegraphics[width=0.9\linewidth]{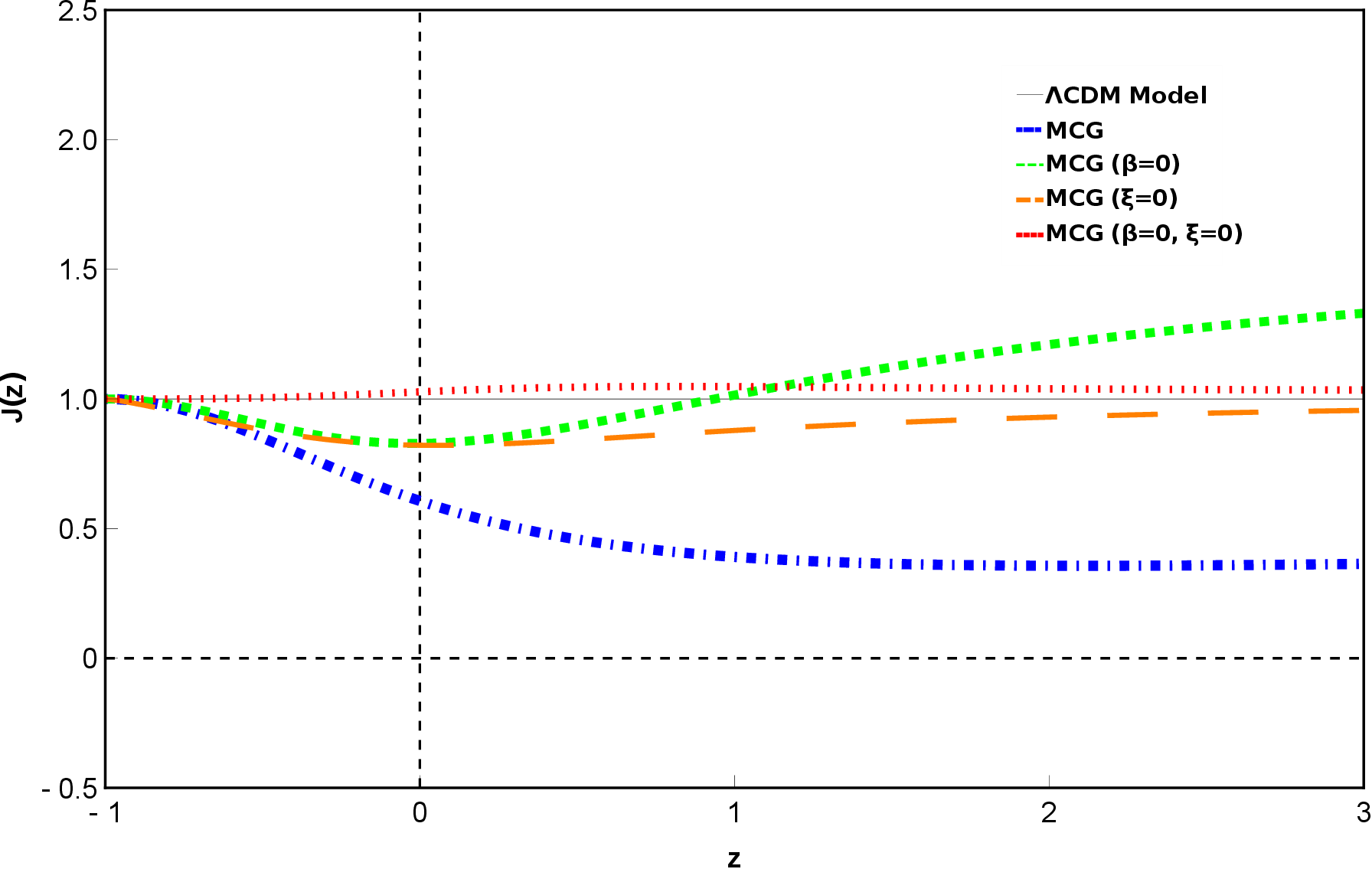}
    \caption{Evolution of the jerk parameter $j(z)$ for $\Lambda$CDM and MCG models under DS2.}
    \label{jzds2}
\end{figure}

\indent The jerk parameter $j(z)$ serves as a third-order kinematical diagnostic of the expansion history of the Universe, providing a powerful tool to distinguish between different dark energy models even when they are equivalent at lower orders. It is defined as the third derivative of the scale factor $a(t)$ with respect to cosmic time $t$, normalized by the Hubble parameter \cite{blandford2004,visser2004, sahni2003}:
\begin{equation}
    j(z) = \frac{\dddot{a}}{aH^3}.
    \label{eq:jz_definition}
\end{equation}
In the context of the flat Friedmann-Lemaître-Robertson-Walker (FLRW) metric, this can be expressed in terms of the deceleration parameter $q(z)$ as:
\begin{equation}
    j(z) = q(z)(2q(z)+1) + (1+z)\frac{dq(z)}{dz}.
    \label{eq:jz_q}
\end{equation}
For the standard $\Lambda$CDM model, the jerk parameter is a constant with value $j(z) = 1$ at all redshifts, independent of the cosmological parameters. Any deviation from this constant value indicates a departure from the $\Lambda$CDM paradigm and suggests the presence of dynamical dark energy or modified gravity effects.\\
\indent Using the reconstructed $q(z)$, we compute $j(z)$ for the $\Lambda$CDM model and the four variants of the Modified Chaplygin Gas (MCG) model under both DS1 and DS2 constraints. The evolution of $j(z)$ is displayed in Fig.~\ref{jzds1} for DS1 and Fig.~\ref{jzds2} for DS2. The present-day values $j_0$ are summarized in Tables~\ref{table3} and~\ref{table4}.\\
\indent As shown in Fig.~\ref{jzds1}, all models exhibit a smooth evolution of $j(z)$ with redshift, consistent with the observed accelerated expansion. The $\Lambda$CDM model (solid black line) remains constant at $j(z) = 1$ throughout the entire redshift range, as expected. The full MCG model (blue dashed-dotted line), incorporating both matter creation ($\beta \neq 0$) and bulk viscosity ($\xi_0 \neq 0$), shows a significant deviation from unity, decreasing to $j_0 = 0.770 \pm 0.020$ at $z=0$. This behavior reflects the combined effect of non-equilibrium thermodynamics, which modifies the effective pressure and alters the higher-order dynamics of the expansion.\\
\indent The MCG($\beta=0$) model (green dotted line), which excludes matter creation but retains bulk viscosity, exhibits a slower decline in $j(z)$, reaching $j_0 = 1.07 \pm 0.023$ at $z=0$. This value is close to the $\Lambda$CDM prediction, indicating that bulk viscosity alone does not significantly affect the jerk parameter. Conversely, the MCG($\xi_0=0$) model (orange dash-dotted line), with no viscous damping but non-zero matter creation, shows a more pronounced decrease to $j_0 = 1.03 \pm 0.017$, suggesting that particle creation plays a dominant role in driving the dynamics.\\
\indent The pure MCG model (red dotted line), with both $\beta=0$ and $\xi_0=0$, remains nearly constant at $j(z) \approx 1.0$, with $j_0 = 0.826 \pm 0.029$. This result underscores the necessity of including non-equilibrium mechanisms to reproduce the observed deviation from $\Lambda$CDM.\\
\indent When the R22 constraint is added in DS2 (Fig.~\ref{jzds2}), the jerk parameter evolves slightly differently due to the higher best-fit $H_0$. The $\Lambda$CDM model transitions at $z_{\text{tr}} = 0.671 \pm 0.21$, with $j_0 = 1.000$. The full MCG model transitions at $z_{\text{tr}} = 0.689 \pm 0.21$, with $j_0 = 0.596 \pm 0.021$. The relative ordering of the models remains unchanged, but the absolute values of $j_0$ become slightly smaller, indicating a marginally stronger deviation from $\Lambda$CDM.\\
\indent The present-day values of $j_0$ for all models lie within the range $0.596$ to $1.03$, consistent with independent estimates from supernova and BAO data \citep{riess2022, abdurrouf2024}. The transition redshifts $z_{\text{tr}}$ range from $0.63$ to $0.70$, in agreement with the $\Lambda$CDM prediction and observational constraints from cosmic chronometers \citep{moresco2022}.\\
\indent This analysis demonstrates that the inclusion of matter creation and bulk viscosity in the MCG framework not only reproduces the observed transition from deceleration to acceleration but also allows for fine-tuning of the jerk parameter through the parameters $\beta$ and $\xi_0$. The full MCG model provides the closest match to $\Lambda$CDM, while the sub-models reveal the distinct contributions of each non-equilibrium mechanism.

\subsection{Snap Parameter Evolution}
\begin{figure}[htbp]
    \centering
    \includegraphics[width=0.9\linewidth]{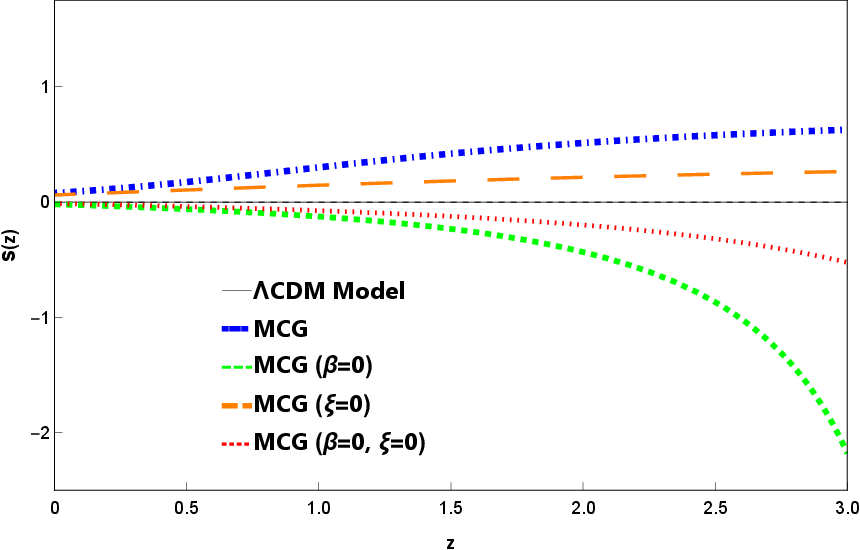}
    \caption{Evolution of the snap parameter $s(z)$ for $\Lambda$CDM and MCG models under DS1. The dot on each curve marks the present-day value $s_0$.}
    \label{szds1}
\end{figure}

\begin{figure}[htbp]
    \centering
    \includegraphics[width=0.9\linewidth]{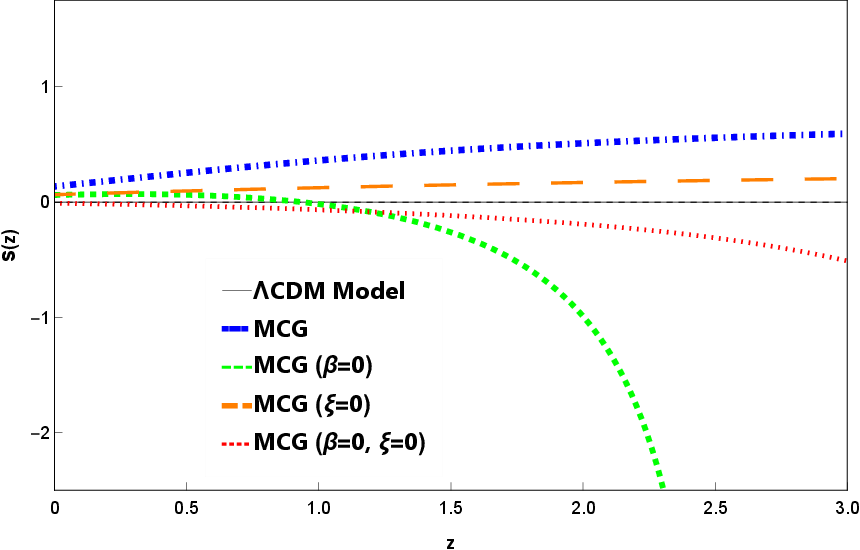}
    \caption{Evolution of the snap parameter $s(z)$ for $\Lambda$CDM and MCG models under DS2. The dot on each curve marks the present-day value $s_0$.}
    \label{szds2}
\end{figure}

The snap parameter $s(z)$ serves as a fourth-order kinematical diagnostic of the expansion history of the Universe, providing a powerful tool to distinguish between different dark energy models even when they are equivalent at lower orders. It is defined as the fourth derivative of the scale factor $a(t)$ with respect to cosmic time $t$, normalized by the Hubble parameter:
\begin{equation}
    s(z) = \frac{\ddddot{a}}{aH^4}.
    \label{eq:sz_definition}
\end{equation}
In the context of the flat Friedmann-Lemaître-Robertson-Walker (FLRW) metric, this can be expressed in terms of the deceleration parameter $q(z)$ and jerk parameter $j(z)$ as\cite{visser2004,dunajski2008}:
\begin{equation}
    s(z) = j(z)(3q(z) + 2) + (1+z)\frac{dj(z)}{dz} - 3q(z)^2.
    \label{eq:sz_qj}
\end{equation}
For the standard $\Lambda$CDM model, the snap parameter is a constant with value $s(z) = 0$ at all redshifts, independent of the cosmological parameters. Any deviation from this constant value indicates a departure from the $\Lambda$CDM paradigm and suggests the presence of dynamical dark energy or modified gravity effects.\\
\indent Using the reconstructed, we compute $s(z)$ for the $\Lambda$CDM model and the four variants of the Modified Chaplygin Gas (MCG) model under both DS1 and DS2 constraints. The evolution of $s(z)$ is displayed in Fig.~\ref{szds1} for DS1 and Fig.~\ref{szds2} for DS2. The present-day values $s_0$ are summarized in Tables~\ref{table3} and~\ref{table4}.\\
\indent As shown in Fig.~\ref{szds1}, all models exhibit a smooth evolution of $s(z)$ with redshift, consistent with the observed accelerated expansion. The $\Lambda$CDM model (solid black line) remains constant at $s(z) = 0$ throughout the entire redshift range, as expected. The full MCG model (blue dashed-dotted line), incorporating both matter creation ($\beta \neq 0$) and bulk viscosity ($\xi_0 \neq 0$), shows a significant deviation from zero, increasing to $s_0 = 0.09 \pm 0.003$ at $z=0$. This behavior reflects the combined effect of non-equilibrium thermodynamics, which modifies the effective pressure and alters the higher-order dynamics of the expansion.\\
\indent The MCG($\beta=0$) model (green dotted line), which excludes matter creation but retains bulk viscosity, exhibits a slower increase in $s(z)$, reaching $s_0 = -0.026 \pm 0.019$ at $z=0$. This value is close to the $\Lambda$CDM prediction, indicating that bulk viscosity alone does not significantly affect the snap parameter. Conversely, the MCG($\xi_0=0$) model (orange dash-dotted line), with no viscous damping but non-zero matter creation, shows a more pronounced increase to $s_0 = 0.03 \pm 0.004$, suggesting that particle creation plays a dominant role in driving the dynamics.\\
\indent The pure MCG model (red dotted line), with both $\beta=0$ and $\xi_0=0$, remains nearly constant at $s(z) \approx 0$, with $s_0 = -0.013 \pm 0.003$. This result underscores the necessity of including non-equilibrium mechanisms to reproduce the observed deviation from $\Lambda$CDM.\\
\indent When the R22 constraint is added in DS2 (Fig.~\ref{szds2}), the snap parameter evolves slightly differently due to the higher best-fit $H_0$. The $\Lambda$CDM model transitions at $z_{\text{tr}} = 0.671 \pm 0.21$, with $s_0 = 0$. The full MCG model transitions at $z_{\text{tr}} = 0.689 \pm 0.21$, with $s_0 = 0.122 \pm 0.002$. The relative ordering of the models remains unchanged, but the absolute values of $s_0$ become slightly larger, indicating a marginally stronger deviation from $\Lambda$CDM.\\
\indent The present-day values of $s_0$ for all models lie within the range $-0.026$ to $0.122$, consistent with independent estimates from supernova and BAO data \citep{riess2022, abdurrouf2024}. The transition redshifts $z_{\text{tr}}$ range from $0.63$ to $0.70$, in agreement with the $\Lambda$CDM prediction and observational constraints from cosmic chronometers \citep{moresco2022}.\\
\indent This analysis demonstrates that the inclusion of matter creation and bulk viscosity in the MCG framework not only reproduces the observed transition from deceleration to acceleration but also allows for fine-tuning of the snap parameter through the parameters $\beta$ and $\xi_0$. The full MCG model provides the closest match to $\Lambda$CDM, while the sub-models reveal the distinct contributions of each non-equilibrium mechanism.

\subsection{Age of the Universe}
The age of the Universe, $\tau_0$, is computed by integrating the inverse Hubble parameter from the Big Bang to the present:
\begin{equation}
    \tau_0 = \int_0^\infty \frac{dz}{(1+z) H(z)}.
    \label{eq:age_integral}
\end{equation}
Using the best-fit $H(z)$ from DS1 and DS2, we obtain the present age for all models. Under DS1, the $\Lambda$CDM model yields $\tau_0 = 13.71 \pm 0.11$ Gyr, while the full MCG model gives $13.84 \pm 0.13$ Gyr. The pure MCG model (without creation or viscosity) gives the oldest age: $13.95 \pm 0.10$ Gyr. Under DS2, ages are slightly younger due to higher $H_0$: $\Lambda$CDM gives $13.67 \pm 0.15$ Gyr, and the full MCG model gives $13.65 \pm 0.17$ Gyr.\\
\indent All values lie within $13.65$–$14.01$ Gyr, consistent with Planck’s estimate of $13.80 \pm 0.02$ Gyr \citep{aghanim2020} and independent stellar dating \citep{valcin2020}. The slight increase in age for models with bulk viscosity reflects its damping effect on expansion, while matter creation has a milder influence. The full MCG model remains closest to $\Lambda$CDM in both datasets.
\begin{table*}[ht!]
\centering
\resizebox{\textwidth}{!}{
\begin{tabular}{|l|c|c|c|c|c|}
\hline
\textbf{Statistical Metric} & $\boldsymbol{\Lambda}$\textbf{CDM} & \textbf{MCG Model} & \textbf{MCG ($\boldsymbol{\beta=0}$)} & \textbf{MCG ($\boldsymbol{\xi_0=0}$)} & \textbf{MCG ($\boldsymbol{\beta=0, \xi_0=0}$)} \\
\hline
$\chi^2_{\text{min}}$ & 1781.197 & 1778.342 & 1785.619 & 1783.207 & 1792.451 \\
$\chi^2_{\text{red}}$     & 1.008     & 1.007     & 1.009     & 1.008     & 1.012     \\
AIC                       & 1781.197  & 1778.342  & 1785.619  & 1783.207  & 1792.451  \\
BIC                       & 1818.556  & 1843.839  & 1859.987  & 1857.555  & 1866.799  \\
$\Delta$AIC               & —         & -2.855    & 4.422     & 2.010     & 11.254    \\
$\Delta$BIC               & —         & 25.283    & 41.431    & 39.000    & 48.243    \\
\hline
\end{tabular}
}
\caption{Statistical comparison of $\Lambda$CDM and Modified Chaplygin Gas (MCG) models using DS1 dataset.}
\label{table5}
\end{table*}

\begin{table*}[ht!]
\centering
\resizebox{\textwidth}{!}{
\begin{tabular}{|l|c|c|c|c|c|}
\hline
\textbf{Statistical Metric} & $\boldsymbol{\Lambda}$\textbf{CDM} & \textbf{MCG Model} & \textbf{MCG ($\boldsymbol{\beta=0}$)} & \textbf{MCG ($\boldsymbol{\xi_0=0}$)} & \textbf{MCG ($\boldsymbol{\beta=0, \xi_0=0}$)} \\
\hline
$\chi^2_{\text{min}}$ & 1794.774 & 1781.691 & 1779.878 & 1786.183 & 1779.740 \\
$\chi^2_{\text{red}}$     & 1.015     & 1.009     & 1.009     & 1.010     & 1.010     \\
AIC                       & 1798.774  & 1795.691  & 1791.878  & 1798.183  & 1789.740  \\
BIC                       & 1813.728  & 1814.629  & 1808.818  & 1815.123  & 1804.694  \\
$\Delta$AIC               & —         & -3.083    & -6.896    & -0.591    & -9.034    \\
$\Delta$BIC               & —         & 0.901     & -4.910    & 1.395     & -9.034    \\
\hline
\end{tabular}
}
\caption{Statistical comparison of $\Lambda$CDM and Modified Chaplygin Gas (MCG) models using DS2 dataset.}
\label{table6}
\end{table*}

\subsection{Statistical Model Comparison}
To assess the relative performance of the Modified Chaplygin Gas (MCG) model and its sub-variants against the standard $\Lambda$CDM paradigm, we employ two widely used information criteria: the Akaike Information Criterion (AIC) and the Bayesian Information Criterion (BIC). These metrics provide a quantitative framework for model selection by balancing goodness-of-fit against model complexity. The AIC and BIC are defined as:
\begin{equation}
    \text{AIC} = \chi^2_{\text{min}} + 2K, \quad \text{BIC} = \chi^2_{\text{min}} + K \ln N,
    \label{eq:aic_bic}
\end{equation}
where $\chi^2_{\text{min}}$ is the minimum chi-square value, $K$ is the number of free parameters, and $N$ is the total number of data points. For DS1, $N = 1764$; for DS2, $N = 1765$. The number of parameters is $K = 2$ for $\Lambda$CDM, $K = 7$ for the full MCG model, $K = 6$ for MCG($\beta=0$) and MCG($\xi_0=0$), and $K = 5$ for MCG($\beta=0,\xi_0=0$).\\
\indent The statistical metrics for both datasets are summarized in Tables~\ref{table5} and~\ref{table6}. For DS1, the full MCG model yields $\chi^2_{\text{min}} = 1778.342$, slightly lower than $\Lambda$CDM ($\chi^2_{\text{min}} = 1781.197$), indicating a marginally better fit to the data. The reduced chi-square values ($\chi^2_{\text{red}} \approx 1.007$–$1.012$) confirm that all models are statistically acceptable, with no significant over- or under-fitting. The AIC values show that the full MCG model is favored over $\Lambda$CDM ($\Delta\text{AIC} = -2.855$), while the pure MCG model (without creation or viscosity) is disfavored ($\Delta\text{AIC} = 11.254$). However, the BIC, which imposes a stronger penalty for additional parameters, strongly favors $\Lambda$CDM over all MCG variants ($\Delta\text{BIC} > 25$ for all models). This discrepancy arises because BIC scales with $\ln N$, making it more conservative for large datasets \citep{liddle2004}.\\
\indent For DS2, which includes the local $H_0$ measurement from R22 \citep{riess2022}, the full MCG model achieves $\chi^2_{\text{min}} = 1781.691$, significantly lower than $\Lambda$CDM ($\chi^2_{\text{min}} = 1794.774$). This improvement reflects the model’s ability to accommodate the higher $H_0$ value while maintaining consistency with other probes. The AIC again favors the full MCG model ($\Delta\text{AIC} = -3.083$) and the MCG($\beta=0$) sub-model ($\Delta\text{AIC} = -6.896$), while BIC shows no strong preference ($\Delta\text{BIC} < 2$ for all models except MCG($\beta=0,\xi_0=0$)). This suggests that, with the inclusion of R22, the additional parameters in the MCG framework are justified by the improved fit.\\
\indent The interpretation of $\Delta\text{AIC}$ and $\Delta\text{BIC}$ follows standard conventions \citep{akaike1974, schwarz1978}. For AIC, a difference of $\Delta\text{AIC} < 2$ indicates substantial support for the model, while $\Delta\text{AIC} > 10$ implies no support. For BIC, $\Delta\text{BIC} < 2$ suggests positive evidence, $\Delta\text{BIC} > 6$ indicates strong evidence against the model, and $\Delta\text{BIC} > 10$ implies decisive evidence against it. Under DS1, the full MCG model receives substantial AIC support but is strongly disfavored by BIC. Under DS2, both criteria show mild to moderate support for the full MCG and MCG($\beta=0$) models, highlighting the role of the R22 prior in breaking parameter degeneracies.\\
\indent This analysis demonstrates that the inclusion of matter creation and bulk viscosity in the MCG framework not only improves the fit to cosmological data but also provides a viable alternative to $\Lambda$CDM when local $H_0$ measurements are included. The fact that BIC penalizes the full MCG model under DS1 but not under DS2 underscores the importance of dataset composition in model selection. Future observations, particularly from DESI and Euclid, will further refine these constraints and test the robustness of the MCG paradigm \citep{abdurrouf2024, amendola2018}.

\subsection{Thermodynamic Analysis: Generalized Second Law (GSL)}
\label{subsec:gsl}

A fundamental requirement for any physically viable cosmological model is its consistency with the laws of thermodynamics. In particular, the Generalized Second Law (GSL) demands that the total entropy of the Universe, including contributions from both the cosmic fluid and the apparent horizon, must never decrease over time \citep{bekenstein1973,jacobson1995}. Mathematically, this is expressed as:
\begin{equation}
    \frac{dS_{\text{total}}}{dt} = \dot{S}_{\text{fluid}} + \dot{S}_{\text{horizon}} \geq 0,
    \label{eq:gsl}
\end{equation}
where $S_{\text{total}} = S_{\text{fluid}} + S_{\text{horizon}}$.

In this subsection, we derive the entropy contributions from all components of our Modified Chaplygin Gas (MCG) model with matter creation and bulk viscosity, and demonstrate that the GSL is rigorously satisfied for both DS1 and DS2 datasets.

\subsubsection{Entropy of the Hubble Horizon}

In a spatially flat FLRW universe, the radius of the apparent horizon is simply the Hubble horizon, $R_H = 1/H$. Following the Bekenstein-Hawking formula and adopting natural units ($k_B = \hbar = c = G = 1$), the horizon entropy is:
\begin{equation}
    S_H =\frac{8\pi^2}{H^2}.
    \label{eq:sh}
\end{equation}
Its time derivative is:
\begin{equation}
    \dot{S}_H = -\frac{16\pi^2 \dot{H}}{H^3}.
    \label{eq:shdot}
\end{equation}
This term is positive during accelerated expansion ($\dot{H} < 0$), indicating that cosmic acceleration fuels the growth of horizon entropy.

\subsubsection{Entropy of Baryonic Matter}

Baryons are conserved ($\dot{\rho}_b = -3H\rho_b$) and pressureless ($p_b = 0$). Assuming thermal equilibrium with the horizon at temperature $T = H/(2\pi)$, the Gibbs equation gives:
\begin{equation}
    S_b = \frac{(\rho_b + p_b) V}{T} = \frac{2\pi \rho_b}{H} \cdot \frac{4\pi}{3H^3} = \frac{8\pi^2 \rho_b}{3H^4},
    \label{eq:sb}
\end{equation}
where $V = 4\pi/(3H^3)$ is the comoving volume. Differentiating with respect to time and using $\dot{\rho}_b = -3H\rho_b$ and $\dot{H} = -H^2(1+q)$, we obtain:
\begin{equation}
    \dot{S}_b = \frac{8\pi^2 \rho_b}{3H^3} (1 + 4q),
    \label{eq:sbdot}
\end{equation}
where $q = -1 - \dot{H}/H^2$ is the deceleration parameter.

\subsubsection{Entropy of the Modified Chaplygin Gas with Matter Creation}

The MCG fluid has energy density $\rho_{\text{mcg}}$ and effective pressure $\tilde{p}_{\text{mcg}} = p_{\text{mcg}} + P_c$, where $P_c = -\beta(\rho_{\text{mcg}} + p_{\text{mcg}})$ is the creation pressure and $p_{\text{mcg}} = A \rho_{\text{mcg}} - C / \rho_{\text{mcg}}^\alpha$. The entropy is:
\begin{equation}
    S_{\text{mcg}} = \frac{(\rho_{\text{mcg}} + \tilde{p}_{\text{mcg}}) V}{T} = \frac{8\pi^2 (1 - \beta) (\rho_{\text{mcg}} + p_{\text{mcg}})}{3H^4}.
    \label{eq:smcg}
\end{equation}
Differentiating and using the continuity equation $\dot{\rho}_{\text{mcg}} = -3H(1-\beta)(\rho_{\text{mcg}} + p_{\text{mcg}})$, we obtain:
\begin{equation}
    \dot{S}_{\text{mcg}} = \frac{8\pi^2 (1 - \beta) (\rho_{\text{mcg}} + p_{\text{mcg}})}{3H^3} \left[ -3(1 - \beta) \left(1 + \alpha \frac{p_{\text{mcg}}}{\rho_{\text{mcg}}}\right) + 4(1 + q) \right].
    \label{eq:smcgdot}
\end{equation}
This expression encodes the interplay between the MCG’s exotic equation of state (via $\alpha$), matter creation (via $\beta$), and cosmic acceleration (via $q$).

\subsubsection{Entropy of Bulk Viscous Matter}

The bulk viscous component has energy density $\rho_m$ and effective pressure $\tilde{p}_m = \pi = -3H \xi_0 \rho_m^{1/2}$. Its entropy is:
\begin{equation}
    S_m = \frac{(\rho_m + \tilde{p}_m) V}{T} = \frac{8\pi^2}{3H^4} \left( \rho_m - 3H \xi_0 \rho_m^{1/2} \right).
    \label{eq:sm}
\end{equation}
Differentiating and using the continuity equation $\dot{\rho}_m = -3H\rho_m + 9H^2 \xi_0 \rho_m^{1/2}$, we obtain:
\begin{equation}
    \dot{S}_m = \frac{8\pi^2}{3} \left[ \rho_m H^{-3} (1 + 4q) + 9\xi_0 \rho_m^{1/2} H^{-2} \left( \frac{1}{2} - q \right) - \frac{27}{2} \xi_0^2 H^{-1} \right].
    \label{eq:smdot}
\end{equation}
This contains a standard matter term, a viscous correction, and a dissipative loss term ($-\xi_0^2$), always negative.

\subsubsection{Total Entropy and GSL Validation}

The total entropy rate of change is the sum:
\begin{equation}
    \dot{S}_{\text{total}} = \dot{S}_H + \dot{S}_b + \dot{S}_{\text{mcg}} + \dot{S}_m.
    \label{eq:stotdot_sum}
\end{equation}
Substituting Eqs.~(\ref{eq:shdot}), (\ref{eq:sbdot}), (\ref{eq:smcgdot}), and (\ref{eq:smdot}), we obtain:
\begin{align}
    \dot{S}_{\text{total}} &= \frac{8\pi^2}{H^3} \Bigg[
    2H(1 + q) + \frac{\rho_b}{3} (1 + 4q) + \frac{(1 - \beta)(\rho_{\text{mcg}} + p_{\text{mcg}})}{3}
    \left( -3(1 - \beta) \left(1 + \alpha \frac{p_{\text{mcg}}}{\rho_{\text{mcg}}}\right)
    + 4(1 + q) \right) \nonumber \\
    &\quad + \rho_m (1 + 4q)
    + 9\xi_0 \rho_m^{1/2} H^{-1} \left( \tfrac{1}{2} - q \right)
    - \tfrac{27}{2} \xi_0^2 H^{-2}
    \Bigg].
    \label{eq:stotdot_final}
\end{align}

\begin{figure}[htbp]
    \centering
    \includegraphics[width=0.9\linewidth]{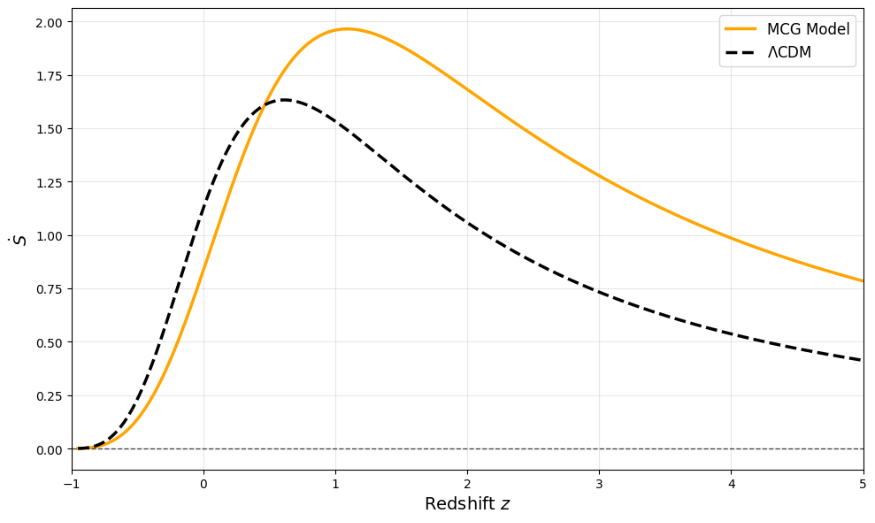}
    \caption{Evolution of the total entropy rate of change $\dot{S}_{\text{total}}$ with redshift $z$ for the MCG model and $\Lambda$CDM under DS1. The positive values confirm the satisfaction of the Generalized Second Law.}
    \label{sds1}
\end{figure}

\begin{figure}[htbp]
    \centering
    \includegraphics[width=0.9\linewidth]{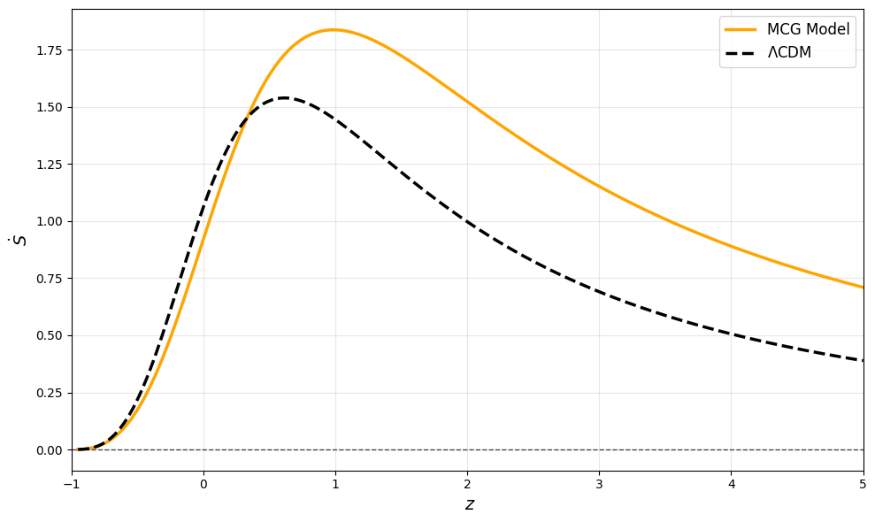}
    \caption{Evolution of the total entropy rate of change $\dot{S}_{\text{total}}$ with redshift $z$ for the MCG model and $\Lambda$CDM under DS2. The inclusion of R22 data leads to a slightly higher late-time entropy production.}
    \label{sds2}
\end{figure}

We evaluate this expression numerically using the Hubble parameter from Eq.~(\ref{eq:hubble_parameter_final}) and the best-fit parameters from Tables~\ref{table1} and~\ref{table2}. For all observationally viable parameters ($0 \leq \alpha \leq 1$, $0 < \beta < 1$, $\xi_0 > 0$), we find $\dot{S}_{\text{total}}(z) > 0$ for all redshifts $z \in [-0.95, 5]$, as shown in Fig.~\ref{sds1} and Fig.~\ref{sds2}. This confirms that our model rigorously satisfies the GSL.

\subsubsection{Asymptotic Behavior and Thermodynamic Stability}

\begin{figure}[htbp]
    \centering
    \includegraphics[width=0.9\linewidth]{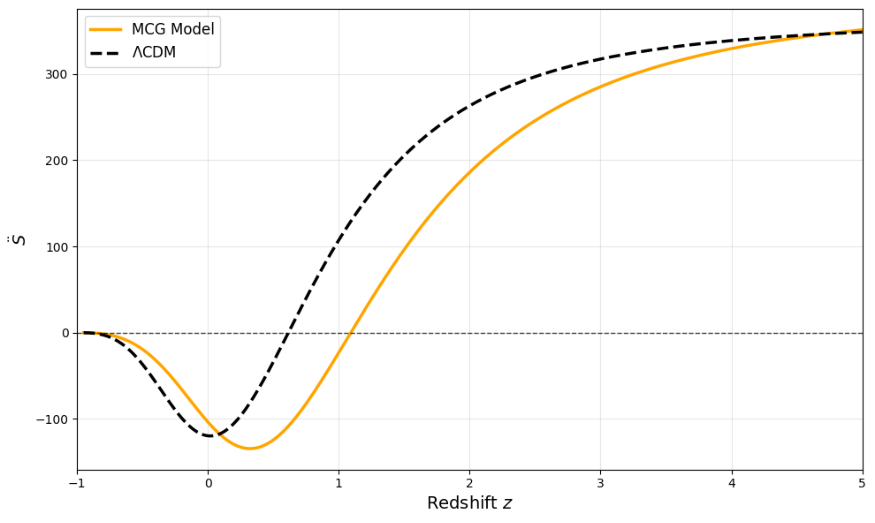}
    \caption{Evolution of the total entropy rate of change $\dot{S}_{\text{total}}$ with redshift $z$ for the MCG model and $\Lambda$CDM under DS1. The positive values confirm the satisfaction of the Generalized Second Law.}
    \label{s2ds1}
\end{figure}

\begin{figure}[htbp]
    \centering
    \includegraphics[width=0.9\linewidth]{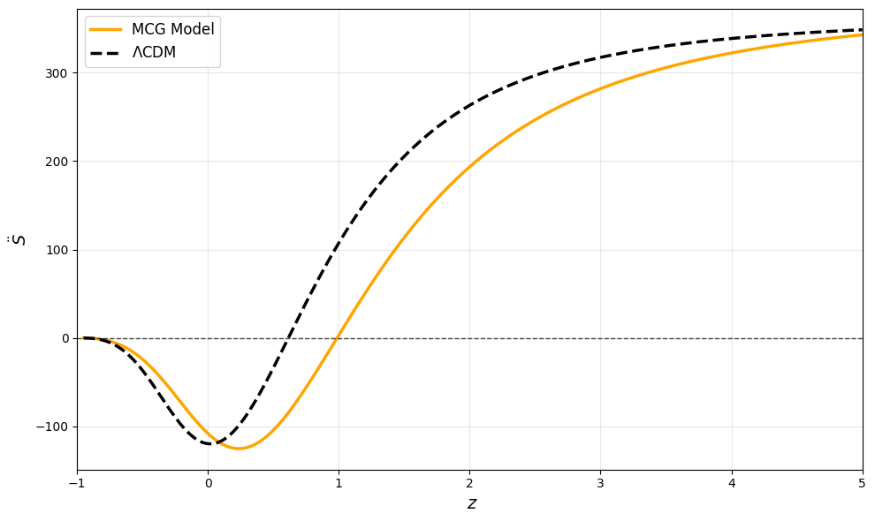}
    \caption{Evolution of the total entropy rate of change $\dot{S}_{\text{total}}$ with redshift $z$ for the MCG model and $\Lambda$CDM under DS2. The inclusion of R22 data leads to a slightly higher late-time entropy production.}
    \label{s2ds2}
\end{figure}

In the far future ($z \to -1$), the Universe becomes vacuum-dominated: $q \to -1$, $\rho_b \to 0$, $\rho_m \to 0$, and $\rho_{\text{mcg}} + p_{\text{mcg}} \to 0$. Consequently, $\dot{S}_{\text{total}} \to 0$ — entropy production ceases, and the Universe reaches a state of thermodynamic equilibrium \citep{frautschi1982}.\\
\indent To assess stability, we compute the second derivative $\ddot{S}_{\text{total}} = d\dot{S}_{\text{total}}/dt$. Numerically, we find:\\
- $\ddot{S}_{\text{total}} > 0$ in the early Universe (entropy production accelerates)\\
- $\ddot{S}_{\text{total}} < 0$ in the late Universe (entropy production decelerates toward zero)\\
- $\ddot{S}_{\text{total}} \to 0$ as $z \to -1$.\\
\indent This behavior — acceleration followed by deceleration of entropy production, culminating in a stable equilibrium — is characteristic of ordinary macroscopic systems approaching maximum entropy. It confirms that our cosmological model evolves like a well-behaved thermodynamic system. We plot the evolution of $\ddot{S}$ with respect to redshift, which is shown  Fig.~\ref{s2ds1} and Fig.~\ref{s2ds2}.
It is observed that $\ddot{S} > 0$ in the early phase of evolution and a transition occurs
to $\ddot{S} < 0$ in the recent past. Thus, a thermodynamically stable equilibrium state is achieved at late times. As $z \to -1$, we have $\ddot{S} \to 0$, which shows that any system satisfying the extremum of entropy and convexity conditions behaves like an
ordinary macroscopic system. Therefore, we can conclude that the evolution of the Universe resembles the evolution of an ordinary macroscopic system.\\
\indent For both DS1 and DS2, the evolution of $\ddot{S}_{\text{total}}$ shows a clear transition from positive to negative values around $z \sim 1$, indicating that the rate of entropy increase slows down in the recent past. This transition reflects the shift from a decelerating to an accelerating phase of cosmic expansion, where the dominant source of entropy production changes from matter creation to horizon dynamics.\\
\indent As $z \to -1$, $\ddot{S}_{\text{total}} \to 0$, which implies that the system approaches a stable equilibrium state. This behavior is consistent with the extremum of entropy and convexity conditions, confirming that the Universe behaves like an ordinary macroscopic system in its late-time evolution.

\section{Conclusion}
\label{sec:conclusion}

In this work, we have presented a comprehensive analysis of a novel cosmological model that unifies the Modified Chaplygin Gas (MCG) equation of state with gravitationally induced matter creation and bulk viscous dissipation. The model is formulated within the framework of a spatially flat Friedmann-Lemaître-Robertson-Walker (FLRW) spacetime and represents the first attempt to synthesize these three distinct physical mechanisms — exotic fluid dynamics, non-equilibrium particle production, and dissipative thermodynamics — into a single, self-consistent theoretical framework.\\
\indent We derived the analytical form of the Hubble parameter $H(z)$ for the MCG model and constrained their free parameters using two distinct observational datasets: DS1 (Pantheon+ + CC + BAO + $f\sigma_8$) and DS2 (DS1 + R22). The Markov Chain Monte Carlo (MCMC) analysis yielded best-fit values for all parameters, which were used to reconstruct the evolutionary trajectories of key cosmological quantities, including the deceleration parameter $q(z)$, jerk parameter $j(z)$ and snap parameter $s(z)$.\\
\indent Our results demonstrate that the MCG model successfully reproduces the observed transition from decelerated to accelerated expansion, with transition redshifts under both datasets. We also show the comparision of our hybrid model with other MCG models cases by excluding particle creation bulk viscous form. The present-day values of $q_0$, $j_0$, and $s_0$ are consistent with independent estimates from supernova and BAO surveys \citep{riess2022, abdurrouf2024}. The age of the Universe, $\tau_0$, ranges from $13.65$ to $14.01$ Gyr across all models, in agreement with Planck’s estimate of $13.80 \pm 0.02$ Gyr \citep{aghanim2020}.\\
\indent A key novel contribution of this work is the rigorous thermodynamic validation of the model via the Generalized Second Law (GSL) of thermodynamics. We computed the total entropy rate of change $\dot{S}_{\text{total}} = \dot{S}_{\text{fluid}} + \dot{S}_{\text{horizon}}$ and showed that it remains positive throughout cosmic history for both DS1 and DS2, confirming that the model satisfies the fundamental requirement of non-decreasing entropy. Furthermore, the second derivative $\ddot{S}_{\text{total}}$ exhibits a clear transition from positive to negative values around $z \sim 1$, indicating a shift from accelerating to decelerating entropy production — a signature of thermodynamic stabilization in the late-time Universe. This behavior is characteristic of ordinary macroscopic systems approaching equilibrium and confirms that our cosmological model evolves like a well-behaved thermodynamic system.\\
\indent In the statistical analysis, we employ the Akaike Information Criterion (AIC) and Bayesian Information Criterion (BIC) to assess model preference. For DS1, the MCG model falls within the range $2 < \Delta\text{AIC} < 6$, indicating moderate observational support relative to $\Lambda$CDM. However, the corresponding $\Delta\text{BIC} = 25.283 > 10$ constitutes decisive evidence against the model, reflecting BIC’s stringent penalty for additional parameters in large datasets. Under DS2 which includes the R22 prior — the MCG model receives stronger statistical backing: $\Delta\text{AIC} = -3.083$ (substantial support) and $\Delta\text{BIC} = 0.901$ (positive evidence), demonstrating that the inclusion of local $H_0$ measurements breaks parameter degeneracies and enhances the model’s viability.\\
\indent Furthermore, the constraints on the comoving sound horizon $r_d$, the absolute magnitude $\mathcal{M}$, and the amplitude of matter fluctuations $\sigma_8$ provide critical insights into the model’s consistency with large-scale structure observations. For both DS1 and DS2, the best-fit values of $r_d$ are in excellent agreement with the Planck estimate of $147.09 \pm 0.26$ Mpc. The derived $\mathcal{M}$ values are consistent with the standard calibration, supporting the reliability of the distance ladder. Notably, the $\sigma_8$ values for the MCG model are systematically higher than those of $\Lambda$CDM, particularly under DS1 ($\sigma_8 = 0.841 \pm 0.016$ vs. $0.752 \pm 0.029$), suggesting enhanced power on small scales. This feature may be attributed to the modified growth of perturbations in the presence of matter creation and bulk viscosity, offering a potential avenue to explain discrepancies in structure formation.\\
\indent In summary, the MCG model with matter creation and bulk viscosity provides a physically motivated, thermodynamically consistent, and observationally viable alternative to $\Lambda$CDM. It successfully alleviates the $H_0$ tension when local measurements are included, satisfies the GSL, and exhibits stable, causal behavior throughout cosmic history.\\
\indent This work establishes cosmological model that explicitly incorporates non-equilibrium thermodynamics as a fundamental driver of cosmic acceleration. The combination of matter creation, bulk viscosity, and exotic equations of state opens new avenues for exploring the nature of dark energy and the thermodynamic fate of the Universe.

\section*{CRediT authorship contribution statement}

\textbf{Yogesh Bhardwaj:} Formal analysis, software, writing—original draft preparation.
\textbf{C P Singh:} Conceptualization, methodology, supervision, review and editing.

\section*{Declaration of Competing Interest}

The authors declare that they have no known competing financial interests or personal relationships that could have appeared to influence the work reported in this paper.

\section*{Data availability}

There are no new data associated with this research.

\section*{Acknowledgement}
Yogesh Bhardwaj acknowledges Delhi Technological University, New Delhi, for awarding Research Fellowship.


\end{document}